\documentstyle[preprint,aps,pre,floats]{revtex}
\tighten
\begin{document}
\input epsf
\begin{figure}
\end{figure}
\begin{table}
\end{table}
\newpage
\setcounter{page}{1}
\preprint{WIS-98-ED}
\draft
\title{ Self-Averaging, Distribution of Pseudo-Critical Temperatures and 
        Finite Size Scaling in Critical Disordered Systems  }
\author{Shai Wiseman and Eytan Domany}
\address{Department of Physics of Complex Systems, Weizmann Institute of
science, \\          Rehovot 76100 Israel }
\date{Submitted to Phys. Rev. E, Feb. 9, 1998}
\newcommand{\av}[1]{\langle #1 \rangle}
\maketitle

\vspace*{-5mm}
\begin{abstract}
\vspace*{-5mm}

We evaluate by Monte Carlo simulations various 
singular thermodynamic quantities $X$, for  an
ensemble of quenched random Ising and Ashkin-Teller models. 
The measurements
are taken at $T_c$ and we study how 
the distributions $P(X)$ (and, in particular, their   
relative squared width, $R_X$)
over the ensemble depend on the system size $l$.
The Ashkin-Teller model was studied in the regime where 
bond randomness is irrelevant and 
we found  weak self averaging;
$R_X\sim l^{\frac{\alpha}{\nu}} \rightarrow 0$, where $\alpha<0$ 
 and $\nu$ are the exponents (of the pure model fixed point)
governing the transition. For the site dilute Ising model on a cubic lattice,
known to be governed by a random fixed point,
we find that $R_X$ tends to a universal constant independent of the amount of
 dilution (no self averaging). 
However this constant is different for canonical and grand 
canonical disorder. We identify the pseudo-critical 
temperatures $T_c(i,l)$ of each sample $i$, as the temperature at which the 
susceptibility reaches its maximal value. The  
distribution of these  
$T_c(i,l)$ over the ensemble was investigated;  we found that its variance 
scales as
$\left(\delta T_c(l)\right)^2 \sim l^{-\frac{2}{\nu}}$.
These results are in agreement with the recent predictions of
Aharony and Harris.  Our previously proposed 
finite size scaling ansatz for disordered systems was tested and found to 
hold. When we fit the data obtained for many samples of different sizes
by a sample-independent form, the resulting scaling function was found 
to be universal and to behave similarly to pure
systems. We did observe that to describe deviations from this universal
function we do need sample-dependent scaling functions. These deviations are, 
however, relatively small and this led us to an interesting side
 result:
sample-to-sample fluctuations of $\chi^{max}$, the susceptibility measured
at
 $T_c(i,l)$, 
 are smaller by a factor of 70 than those
of $\chi(T_c)$. 
This indicates that to obtain a fixed statistical error
it may be more computationally efficient to measure $\chi^{max}$.

\end{abstract}



\section{introduction}

 How is the critical behavior affected by the introduction of disorder
(usually dilution or bond--randomness) into a model? This question has been
extensively studied\cite{Stinch} experimentally, analytically\cite{Shalaev94}
 and numerically\cite{Sel Sing} for quite some time now.
 The Harris criterion\cite{Harris,Stinch,Kinz:Dom}
 states that $\phi$, the scaling index of the operator
corresponding to randomness at the pure system fixed point
(also called the crossover exponent ) is equal to
$\frac{\alpha_{\text{p}}}{\nu_{\text{p}}}$ ($ \alpha_{\text{p}}$ and $\nu_{\text{p}}$ are the specific-heat and
correlation length exponents of the pure model). Thus the critical behavior
of the pure system (p) is unaltered by disorder if $\alpha_{\text{p}}<0$
and a weakly disordered system will have the same critical exponents as the
pure one. If $\alpha_{\text{p}}>0$ even a weakly disordered system will not belong to 
the same universality class as the pure one. If $\alpha_{\text{p}}=0$ the situation is
 marginal.

In Renormalization Group (RG) calculations it is possible to determine to which
fixed point a certain disordered model flows and to determine the nature of
this fixed point. A disordered fixed point is
characterized by a fixed distribution (of finite width) of couplings while a
 pure fixed point
is characterized by a delta function type distribution: a single coupling set.
On the other hand, to the best of our knowledge, in the current Monte Carlo
 state of the art, there is no method that can
check directly whether a certain model is governed by a pure or disordered
fixed point.
 Most Monte Carlo studies 
concentrated on calculating critical exponents of a certain disordered model.
 If these exponents agreed
with an RG calculation, this served as indirect evidence to the nature of
the governing fixed point. In such numerical and experimental 
studies of disordered systems near their 
critical point  finite samples with quenched 
disorder are used; any sample $i$ is a particular random realization of the 
quenched disorder. A measurement (or calculation) of any {\em density } of 
an extensive thermodynamic property $X$ (e.g. $X=E,M,C_{h}$ or $\chi$)
 would yield a different value $X_i$ for every sample $i$ because of the 
differences in the realizations of the quenched disorder.
Here $X_i$ is the exact thermal average of the sample, which in an experiment
or a numerical study can only be estimated by $\overline{X_i}$.
 In an ensemble
of  disordered samples of linear size $l$ the values of $X_i$ are governed by
a probability distribution $P(X)$. In most MC studies only the ensemble 
average $[X]$ is studied. As is shown in this study, it is possible to obtain
direct evidence, by MC, as to the nature of the governing fixed point by
studying $P(X)$ and the factors which govern its shape. 
As it turns out, this can be done by studying
the question of {\em self averaging}, which concerns
the behavior of the width of $P(X)$ as the system size is increased. 
 We characterize $P(X)$ by the ensemble average $[X]$ and variance
\begin{equation}
 V_X= [X^2]- [X]^2  
\;. \label{eq: sigma X} \end{equation}

Suppose that $X$ is a singular density  of an extensive
 thermodynamic property, such as $M$, $\chi$ or the singular part of $E$ or
 $C$. The system is said to exhibit {\em self-averaging}\cite{Rev:Mod} 
 if 
\begin{equation}
R_X(l)= V_X / [X]^2 \rightarrow 0\;\;\;\; 
                           \mbox{as}\;\;\;\; l\rightarrow \infty
\;, \label{eq: self averaging} \end{equation}
otherwise, if $R_X$ tends to a constant different from zero, it is said to
 exhibit {\em lack of self averaging}.
In a self averaging system a single very large sample is a sufficient
representative of the ensemble. But in a non self-averaging system a
measurement performed on a
single sample, no matter how large, does not give a meaningful result and
therefore must be repeated using many samples from the ensemble. 
 In a MC study of a self averaging disordered system the number of samples
 needed to obtain
$[X]$ to a given relative accuracy decreases with increasing $l$. On the
other hand, in a non self averaging system the number of samples
which must be simulated is independent of $l$ and the total amount of work
rises very strongly with $l$.

Off criticality, where $l$ is much larger than the correlation length $\xi$,
 as first argued by Brout\cite{Brout}, we may divide the sample $i$ into $n$
 large subsamples (much larger than $\xi$). If
we assume that the coupling between neighboring subsystems is negligible,
then the value of any density of an extensive quantity over the whole sample
$X_i$ is equal to the average of the (independent) values of this quantity over
the subsamples.
Provided the probability distribution of the $X$'s of the subsamples has a
finite variance, then according to the Central Limit theorem the value of
$X_i$ is distributed with a Gaussian probability
distribution around its mean $[X_i]$. The square of the width
of the
Gaussian, $V_{X}$, is proportional to $\frac{1}{n}\sim l^{-d}$.
In this case (\ref{eq: self averaging}) is fulfilled, and $X$ is called 
self-averaging. In such a case, where $R_X\sim l^{-d}$, $X$ is called 
{\em strongly self averaging}\cite{BH:book}. 

Close to criticality, where $\xi\sim l$, the Brout argument does not hold, since
the average of $X$ over neighboring subsamples may not be considered as
 independent. Thus at criticality there is no reason to expect that
 $R_X\sim l^{-d}$. In a previous study\cite{WD Self-Av} we considered the
 question of self-averaging at the critical temperature of the infinite 
lattice, $T_c^\infty$, through Monte Carlo simulations of
 several
 random-bond Ashkin Teller models. Our findings, which are summarized shortly
 below, prompted us to suggest 
 a heuristic finite size scaling theory for disordered systems, based on
 physical considerations similar to those leading to the Harris criterion. 
We characterized every sample $i$ of size $l$ by a pseudo-critical 
temperature $T_c(i,l)$ and introduced the reduced temperature of each sample 
$i$
\begin{equation}
\dot{t}_i= \frac{T-T_c(i,l)}{T_c}
\;. \label{eq: t_dot} \end{equation}
$T_c(i,l)$ was assumed to fluctuate around its 
ensemble average $T_c(l)$ with width $\delta T_c(l)$. 
We then assumed [for samples with $T$ close to $T_c(i,l)$]
a sample dependent finite size scaling form
\begin{equation}
X_i(T,l)= l^\rho \tilde{Q}_i( \dot{t}_i l^{y_t} )
\;. \label{eq: disorder fss} \end{equation}
Here $\rho$ is the exponent characterizing the behavior of $[X]$ at $T_c$;
e.g. for $X=\chi$, $\rho=\frac{\gamma}{\nu}$ .
The form of the function $\tilde{Q}_i(Z)$ (or its coefficients) was assumed
to be sample dependent but the critical exponents $\rho$ and 
$y_t=\frac{1}{\nu}$ were assumed to be universal or sample independent.
Relying on (\ref{eq: disorder fss}) we then related
$R_X$ at $T_c^\infty$ to $\delta T_c(l)$. By assuming that
\begin{equation}
\left(\delta T_c(l)\right)^2 \sim  l^{-d}
\label{eq: delta Tc} \end{equation}
the theory predicted that    when the
specific heat exponent of the disordered model $\alpha<0$ then,
at criticality, 
\begin{equation}
R_X \sim l^{\frac{\alpha}{\nu}}
\;.\label{eq: R scaling WD} \end{equation}
Thus, if $\frac{\alpha}{\nu}=0$, $X$ is non self-averaging, but if
 $ -d < \frac{\alpha}{\nu} < 0$, $X$ is called 
{\em weakly self averaging}\cite{BH:book}.
Note that according to \onlinecite{Chayes 1986} $\frac{\alpha}{\nu}\leq 0$ in 
any disordered system, though claims to the contrary exist 
(e.g. \cite{Singh 1988} and \cite{Pazmandi 1997}).  

In a subsequent study, Aharony and Harris\cite{Aharony 1996} (AH) used 
renormalization group analysis to study the dependence of $P(X)$ on $l$ and
 $\xi$. For $l\gg\xi$ they recovered strong self-averaging:
P(X) approaches a Gaussian with relative variance $R_X\sim(l/\xi)^{-d}$.
For $l\ll\xi$ they found two different behaviors. When randomness was
irrelevant and the system was governed by a pure fixed point they found
\begin{equation}
R_X\sim l^{\left( \frac{\alpha}{\nu} \right)_{\text{p}}} 
\;.\label{eq: R scaling AH pure} \end{equation}
In this case the critical exponents of the disordered model are the same as 
those of a pure model, $\frac{\alpha}{\nu}=
\left( \frac{\alpha}{\nu} \right)_{\text{p}}$, so that 
(\ref{eq: R scaling AH pure}) is in agreement with (\ref{eq: R scaling WD}).
On the other hand, when the system is governed by a random fixed point,
they found that $P(X)$ approaches a universal, $l$ independent shape, and 
$R_X\rightarrow \mbox{const}$ as $l \rightarrow \infty$, which implies
 lack of self-averaging. When $\alpha$ of the random model is negative,
this prediction is in disagreement with (\ref{eq: R scaling WD}).
As AH point out, this disagreement 
between the RG result and our scaling ansatz can be reconciled if we assume
that
in disordered systems governed by a random fixed point, (\ref{eq: delta Tc})
is substituted by
\begin{equation}
\left(\delta T_c(l)\right)^2 \sim  l^{-\frac{2}{\nu}}
\;.\label{eq: delta Tc AH} \end{equation}

 In the Monte Carlo study of \onlinecite{WD Self-Av} several 
bond-disordered Ashkin-Teller models on a square lattice were
simulated. These included  the bond-disordered Ising model where
$\left( \frac{\alpha}{\nu} \right)_{\text{p}}=0_+$ ($C\sim \log l$ at
 $T_c^\infty$), and several bond-disordered Ashkin-Teller models where
 $\left( \frac{\alpha}{\nu} \right)_{\text{p}}>0$. 
According to renormalization group expansions\cite{DD:bax,Shalaev94,Cardy priv}
 around the pure Ising model
these models are governed by the pure Ising fixed point, while according to 
MC results\cite{WD AT}  the  bond-disordered Ashkin-Teller models may be
governed by a different (possibly random) fixed point with $ \frac{\alpha}{\nu} =0_+$. 
 It was found that far from criticality all thermodynamic
quantities which were examined (energy, magnetization, specific heat,
susceptibility) are strongly self averaging,
that is $R_{X}\sim l^{-d}$.
At criticality though, the results indicated that the magnetization $m$ and
the susceptibility $\chi$ are non self averaging.
 The energy $E$ at 
criticality was clearly weakly self
averaging, that is $V_{E}\sim l^{-y_{v}}$ with $0<y_{v}<d$
(Here $E$ includes the analytic and singular parts of the energy).
 The theory (\ref{eq: R scaling WD}) is not applicable in
the asymptotic limit ($l\rightarrow \infty$) to the bond-disordered
Ashkin-Teller model where
$\frac{\alpha}{\nu}=0_{+}$. Nonetheless in the accessible range of lattice
sizes good agreement between the theory and the data for
$V_{\chi}$ and $V_{E}$ was found. In particular $R_\chi$ was shown to behave
very similarly to the specific heat $C$, as suggested by 
(\ref{eq: R scaling WD}), for a wide range of lattice sizes
and for different degrees of disorder. In a very recent MC study of
a mean field Potts glass a lack of self averaging of the order parameter
was found as well\cite{Dillmann 98}. 

In this paper we set out to resolve three issues, neither of which could be 
investigated by studying the Ashkin Teller model at $\alpha_{\text{p}}\geq 0$.
\begin{enumerate}

\item So far the prediction (\ref{eq: R scaling WD}) has not been tested
 numerically for the case $\alpha_{\text{p}} < 0$. In this case randomness is
 irrelevant, $\alpha=\alpha_{\text{p}}$ and the AH prediction 
(\ref{eq: R scaling AH pure}) coincides with (\ref{eq: R scaling WD}).

\item When $\alpha_{\text{p}} > 0$ but at the random fixed point $\alpha < 0$,
the scaling theory (\ref{eq: R scaling WD}) predicts weak self averaging,
in disagreement with the AH result $R_X \rightarrow \mbox{const}$. Resolution
of this discrepancy will shed light on whether the ansatz (\ref{eq: delta Tc})
or the AH prediction (\ref{eq: delta Tc AH}) governs the width of the 
distribution of the pseudocritical temperatures $\delta T_c(l)$.

\item Testing the scaling form (\ref{eq: disorder fss}) :
      We wish to determine whether it holds and whether the sample dependence
enters only via $\dot{t}_i$ or also through the scaling function 
$\tilde{Q}_i$. 
\end{enumerate} 
%
%
With these three goals in mind we set out to examine by Monte Carlo
 simulations 
the question of self averaging at criticality in two different models.
 The first is a bond-disordered Ashkin-Teller model at a point where
$\left( \frac{\alpha}{\nu} \right)_{\text{p}}\approx -0.54 $ (a large negative value
was chosen to yield unambiguous results).
To address the second, more important issue, we simulated
the  site-dilute Ising model on a cubic lattice. 
Because the critical exponent $\alpha_{\text{p}}$ is positive for the three
 dimensional pure Ising model, $\alpha_{\text{p}}\approx 0.11$\cite{Fisher74},
 the Harris criterion
 predicts that randomness will lead to a different critical behavior. 
The disordered model is believed\cite{Mayer 89,Janssen 95} to be governed by
 a random fixed point
with $\alpha<0$. Thus, according to AH, this model should exhibit lack of self 
averaging while according to our finite size scaling theory it should 
exhibit weak self averaging as in (\ref{eq: R scaling WD}), if the assumption
 (\ref{eq: delta Tc}) is valid.

Besides calculating $R_X$ at the 
critical point, 
 we decided to test directly which one of 
(\ref{eq: delta Tc}) or (\ref{eq: delta Tc AH}) is correct.
To this end we calculated the pseudo-critical
temperature of each sample taken from an ensemble of site-diluted 3D Ising
 samples at different lattice sizes. The pseudo-critical temperature was 
defined as the temperature of the maxima of the susceptibility of that sample.
This calculation was done by using the histogram reweighting 
method\cite{Ferrenberg 88,Ferrenberg 91,Munger 91,Swendsen 93,Ferrenberg 95}. 
This method 
allows to use the results of a simulation at one temperature for calculating
the value of thermodynamic observables in a {\em whole range} of nearby
temperatures. We thus obtained for each lattice size distributions
of pseudo-critical temperatures $T_c(i,l)$ and were able to calculate their
mean, $T_c(l)$, and variance, $(\delta T_c(l))^2$.

To investigate the extent to which the scaling form (\ref{eq: disorder fss})
holds, we studied the relationship between
the sample dependent magnetization $m_i(T_c)$ and $T_c(i,l)$ using the data
 collapse technique. We did find  convincing support for the finite size 
scaling ansatz (\ref{eq: disorder fss}) but also found evidence for sample
dependence of the scaling function .

This work is organized as follows. In the first part of Sec. \ref{sec: RBAT}
we define the random bond Ashkin-Teller model which was simulated and
summarize its critical behavior as found by the finite size scaling results. 
In the second
part of Sec. \ref{sec: RBAT} we give our results concerning self-averaging at 
criticality. The results indicate clearly that $R_X$ is weakly self averaging
and are in good agreement with (\ref{eq: R scaling AH pure}).
In Sec. \ref{sec: Ising} we summarize some relevant properties of the site 
dilute Ising
model on a cubic lattice and give some details of the simulation.
Finite size scaling results for some observables at criticality are given
as well. In section \ref{sec: self Ising} we analyze and discuss our results 
concerning self averaging at $T_C^\infty$. These results seem to indicate
the correctness of the AH scenario, whereby $R_X$ is non self averaging.
In Sec. \ref{sec: Tc(i,l)} we study the distributions of the pseudo-critical
temperatures of the site dilute Ising model.
The scaling of $(\delta T_c(l) )^2$ does not agree with (\ref{eq: delta Tc})
but seems to agree with (\ref{eq: delta Tc AH}), giving additional evidence
for the validity of the AH scenario. 
In Sec. \ref{sec: Tc(i,l)} we also analyze the distributions of
the maximal susceptibilities $\chi(T_c(i,l))$, and investigate the 
extent to which the scaling form (\ref{eq: disorder fss}) holds.
 The work is summarized in Sec. \ref{sec: sum}.

\section{Weak self averaging in an Ashkin-Teller model 
                              with irrelevant disorder}
\label{sec: RBAT} 

\subsection{Definition of the model}

The model we study is the random-bond Ashkin-Teller model on a square 
lattice. On every site of the lattice two Ising spin variables, $\sigma_i$
and $\tau_i$, are placed. Denoting by $\av{ij}$ a pair of nearest
neighbor sites, the Hamiltonian is given by
\begin{equation}
{\cal H}= - \sum_{\av{ij}} [ K_{i,j} ( \sigma_i \sigma_j + \tau_i \tau_j )
                         + \Lambda_{i,j} \sigma_i \tau_i \sigma_j \tau_j ]
\;. \label{eq: Ham RBAT} \end{equation}
$K_{i,j}$ and $\Lambda_{i,j}$ are chosen according to
   \begin{equation}
(K_{i,j},\Lambda_{i,j})=\left\{ \begin{array}{ll}
 \,(K^{1},\Lambda^{1}) & \mbox{with probability } \frac{1}{2} \\
 \,(K^{2},\Lambda^{2}) & \mbox{with probability } \frac{1}{2} \end{array}
\right. \label{eq:Ran} \;.   \end{equation}

The homogeneous (or pure) 
model\cite{Ashkin 43} $[(K^1,\Lambda^1)=(K^2,\Lambda^2) \equiv (K,\Lambda)]$
possesses a line of critical points along which critical exponents vary 
continuously, so that it flows under Renormalization Group (RG) onto a line
of fixed points. The 
scaling exponent corresponding to randomness, $\phi= (\alpha/\nu)$, 
which is analytically known\cite{exact,Nienhuis}, also
varies continuously along this line. However along the
part of this line ( $\Lambda \geq 0$) interpolating 
between the Ising ($\Lambda = 0$) and four state Potts ($\Lambda = K$) models 
it takes positive values,
($1 \geq \phi \geq 0$), so that randomness is relevant. Indeed the 
critical behavior of the disordered 
model\cite{Cardy priv,DD:bax,Shalaev94,WD AT} was found to be 
different from that of the pure one. 
The self averaging properties of the disordered model were examined
in \onlinecite{WD Self-Av}, and as discussed in the introduction, a lack of 
self averaging was found. 
Along the other part of this line
( $\Lambda < 0$)  $\phi=\frac{\alpha}{\nu}$  is negative, so that 
according to 
the Harris criterion randomness is irrelevant. Thus a slightly disordered 
model will exhibit the same behavior as its pure version. It turns out, 
according to this study (in agreement with RG 
calculations\cite{Shalaev94,Cardy priv}) 
that a model with finite disorder will flow under RG onto a pure model 
 that is along the part of the line of fixed
points where $\phi=\frac{\alpha}{\nu}<0$.  Thus according to AH
as well as according to (\ref{eq: R scaling WD}) we expect to find weak 
self-averaging at criticality for disordered versions of models corresponding
to this part of the Ashkin Teller critical line ($\Lambda<0$).

The Ashkin Teller model is a convenient paradigm for studying 
critical behavior of disordered systems both because its scaling exponent
corresponding to randomness $\phi=\frac{\alpha}{\nu}$ varies continuously
and because part of the critical manifold of the random-bond model can
be found exactly through duality. In parts of the coupling space
where only two phases exist the self dual manifold 
\begin{equation}
(K^2,\Lambda^2)=(\widetilde{K^{1}},\widetilde{\Lambda^{1}})
\label{eq: self dual}\end{equation}
 is critical.
Here $(\widetilde{K^{1}},\widetilde{\Lambda^{1}})$ are the dual couplings
of $(K^1,\Lambda^1)$ according to the duality transformation of the 
Ashkin-Teller model; a discussion of this point can be found in 
\onlinecite{WD AT}. Since the extent of deviation from pure behavior is 
obviously determined by the difference between the two sets of couplings,
we have chosen to study a model with the ratio
\begin{equation}
\frac{K^2}{K^1}= \frac{1}{10} \;,
\label{eq: K ratio}\end{equation}
so that randomness will be pronounced. In addition, a ratio of
\begin{equation}
\frac{\Lambda^1}{K^1}= -f=-\frac{9}{10} 
\label{eq: f ratio}\end{equation}
was chosen. 
Since $\frac{\alpha}{\nu}$ decreases as $f$ increases and 
$\frac{\alpha}{\nu}=0$ for $f=0$ ( Ising model), we have chosen 
$f=\frac{9}{10}$ so that $\frac{\alpha}{\nu}$ will be a pronounced negative 
number.  
 Equations (\ref{eq: self dual}-\ref{eq: f ratio}) define the couplings
of the model simulated, where the temperature $T$ was absorbed into the 
couplings $K_{ij},\Lambda{ij}$.

The critical behavior of the disordered model is
 compared with that of an {\em Anisotropic Ashkin-Teller} model which is 
used as a
reference pure model\cite{WD AT}.  This model has the same Hamiltonian 
(\ref{eq: Ham RBAT}) but with the couplings  distributed as follows:
   \begin{equation}
(K_{i,j},\Lambda_{i,j})=\left\{ \begin{array}{ll}
 \,(K^{1},\Lambda^{1}) & \mbox{ for bonds $(i,j)$ in the
horizontal direction } \\
 \,(K^{2},\Lambda^{2}) & \mbox{for bonds $(i,j)$ in the vertical direction }
\end{array} \right. \label{eq:Ani} \;.   \end{equation}

In the Monte Carlo simulations we used a cluster algorithm\cite{Wang90a} which is described
 in \onlinecite{WD AT}. The main idea is to embed into the Ashkin-Teller model
an Ising model and simulate it using the Wolff\cite{Wol:1C} single cluster
 algorithm for the Ising model. The number of samples simulated was $n=2000$ for linear
lattice sizes $4\leq l \leq 64$, $n(128)=1200$ for $l=128$, and $n(256)=436$.
For each sample $i$ Monte Carlo estimates of various observables 
$\overline{X_i}$  and their
 errors $\delta \overline{X_i}$ were calculated.
 Next we list results concerning the critical behavior of the
estimated bond-disordered ensemble averages $[\overline{X_i}]$ as a function of
lattice size.

\subsection{ Critical behavior of the model}

Here we give the finite size dependences of averages (over all samples) of 
various observables,
defined as in \onlinecite{WD AT}, at the critical point $T_c^\infty$ defined
 through (\ref{eq: self dual}-\ref{eq: f ratio}).

\begin{figure}
\centering
\epsfxsize=87mm
\centerline{ \epsffile{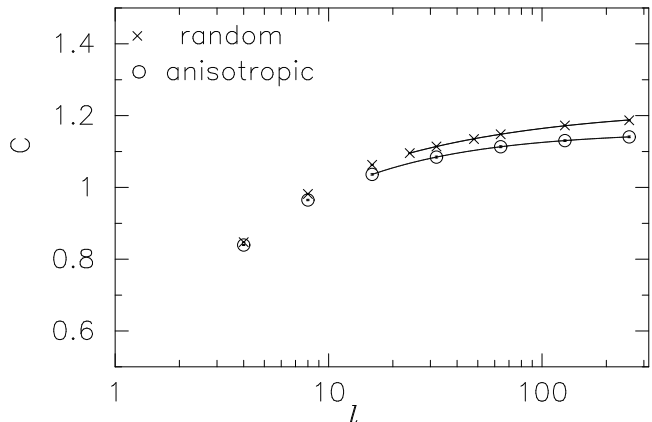} }
\caption{ The specific heat C of the random bond and the anisotropic 
Ashkin Teller models
as a function of $\log l$.
The solid lines are fits to the form (\protect\ref{eq: C}).
 }
 \label{fig: Cv_rc8_} \end{figure}

In figure \ref{fig: Cv_rc8_} we plot the specific heat of the random bond and
 the anisotropic
models as a function of $\log l$. The solid lines are fits to the finite size
scaling form
\begin{equation}
C= B_0 + B_1 l^{\frac{\alpha}{\nu}}
\;.\label{eq: C} \end{equation}
 Using lattice sizes of $16\leq l \leq 256$, we find 
$\frac{\alpha}{\nu}=-.745(4)$ for the anisotropic model, while  using lattice
 sizes of $24\leq l \leq 256$ , we find $\frac{\alpha}{\nu}=-.536(32)$
  for the random bond model (note: $B_1$ is negative). Thus this strongly 
disordered model apparently flows under RG onto a pure model with
different exponents than its anisotropic version but still one that is along
 the part of the line of fixed points where 
$\left(\frac{\alpha}{\nu}\right)_{\text{p}}<0$.

For both models
the magnetization $m$ and susceptibility $\chi$  were fitted to the
 forms
\begin{equation}
m=  A_{m} l^{-\frac{\beta}{\nu}} \;\;\;\;\;\,
\chi= A_\chi l^{\frac{\gamma}{\nu}}
\;.\label{eq: m and Chi } \end{equation}
Similarly the polarization $P$ (magnetization of the $\tau$ spins) and
susceptibility of the polarization $\chi^{(p)}$ were fitted to the
 forms
\begin{equation}
P=  A_{P} l^{-\frac{\beta^{(p)}}{\nu}} \;\;\;\;\;\,
\chi^{(p)}= A_{\chi^{(p)}} l^{\frac{\gamma^{(p)}}{\nu}}
\;.\label{eq: P and Chi-P } \end{equation}
The estimates for the exponent ratios obtained are listed in Table 
\ref{tab: rc8 exponents}.
The values of $\frac{\beta}{\nu}$ and $\frac{\gamma}{\nu}$ are within errors,
for the random bond model or very close, for the anisotropic model, to the
 Ising exponent ratios $\frac{\beta}{\nu}=\frac{1}{8}$ and 
$\frac{\gamma}{\nu}=\frac{7}{4}$. These exponent ratios are predicted 
analytically\cite{exact,Nienhuis} to be of this magnitude all along the  
critical line of the pure Ashkin Teller model.
 
 \begin {table}
\caption{  Estimates for critical exponent ratios of the Ashkin Teller
models from finite size scaling at
 $T_c^\infty$. obtained with linear fits to $\log l$ according to equations
(\protect\ref{eq: C} - \protect\ref{eq: P and Chi-P }).   
}
 \label{tab: rc8 exponents}
\begin{tabular}{lllllll}
   & $\frac{\alpha}{\nu}$& $\frac{\beta}{\nu}$& $\frac{\gamma}{\nu}$& $\frac{\beta^{(p)}}{\nu}$&
                          $\frac{\gamma^{(p)}}{\nu}$& fitting interval \\
\hline
 random bond  & -0.536(32)& 0.1252(4)& 1.7502(5) & 0.322(1) & 1.3566(15)& $24\leq l \leq
 256$         \\
 anisotropic  & -0.745(4)& 0.1262(2)& 1.7488(1) & 0.3312(5)& 1.339(1)  & $16\leq l \leq
 256$\\  
 \end{tabular}
\end{table}


\subsection{Calculation of $V_X$}
\label{sec: calc V_X}
The variance $\sigma^2_X$ of the Monte Carlo estimates $\overline{X_i}$ is the sum of two
 contributions.
The main contribution is due to the variance $V_X$ of the distribution of the 
true $X_i$. $V_X$ is the quantity we wish to study. The second
 contribution is due to the errors of the estimated observables, 
$\delta \overline{X_i}$. Thus, the unbiased estimator\cite{stat1} of the
 variance of the $X_i$ is
\begin{equation}
V_X= \sigma^2_X -[ ( \delta \overline{X_i})^2 ]
\;.\label{eq: V_X def} \end{equation}
$\delta \overline{X_i}$ depends on the length of the MC runs and on the
 autocorrelation time $\tau_X$ of the MC dynamics. To obtain a valid estimate
of $V_X$, $[ ( \delta \overline{X_i})^2 ]/V_X$ should be sufficiently small.
In the random bond Ashkin Teller model studied here this requirement was not
met for the specific heat $C$, whereby we could not study $R_C$. Additional 
discussion of the practical implications of (\ref{eq: V_X def}) can be found
 in section III of \onlinecite{WD Self-Av}. Next we list results concerning 
the critical behavior of $V_X$ at $T_c^{\infty}$. 


\subsection{ The relative variance $R_X$}

\begin{figure}
\centering
\epsfxsize=87mm
\centerline{ \epsffile{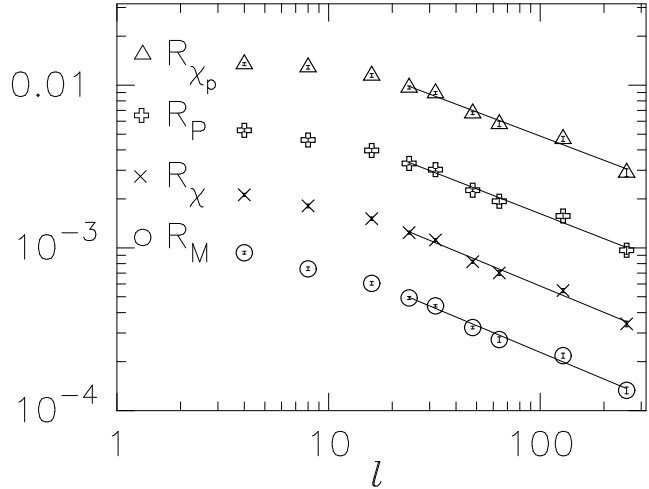} }
\caption{ The relative variance of the susceptibility $R_\chi$, of the
 magnetization $R_m$, of the polarization $R_P$ and of the polarization
 susceptibility $R_{\chi^{(p)}}$  of the Ashkin Teller model at $T_c^{\infty}$
 as a function of
 $\log_{10}l$. The solid lines are linear fits according to 
(\protect\ref{eq: rho-X fit}).
 }
 \label{fig: RXt_rc8} \end{figure}

In Fig. \ref{fig: RXt_rc8} we plot $R_m$, $R_\chi$, $R_P$ and $R_{\chi_p}$
as a function of $\log l$. The solid lines are linear fits to the form 
\begin{equation}
R_X= A_X l^{\rho_{X}}
\label{eq: rho-X fit} \end{equation}
for $24\leq l \leq 256$. The estimates obtained for $\rho_X$ are 
$\rho_\chi=-0.537(32)$, $\rho_m=-0.546(38)$, $\rho_{\chi^{(p)}}=-0.493(37)$ and
$\rho_P=-0.509(41)$. Clearly all observables, $m, \chi, P, \chi^{(p)}$ are
weakly self averaging. Furthermore $\rho_\chi$ and $\rho_m$ are in very good 
agreement with the value of $\frac{\alpha}{\nu}=-.536(32)$, while
 $\rho_{\chi_p}$ and $\rho_P$ are
also within errors of $\frac{\alpha}{\nu}$. Thus the results for the scaling
of the relative variance of these four
observables are in good agreement with the predictions of AH and with
(\ref{eq: R scaling WD}).

\begin{figure}
\centering
\epsfxsize=87mm
\centerline{ \epsffile{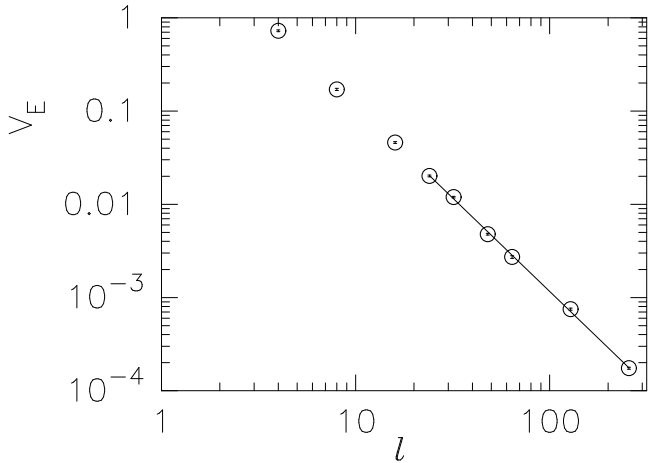} }
\caption{ The variance of the energy $V_E$ of the Ashkin Teller model
  at $T_c^{\infty}$ as a function of
 $\log_{10}l$. The solid line is a linear fit according to 
$V_E= A_{vE} l^{x_E}$, yielding $x_E= -2.005(26)$.
 }
 \label{fig: VE_rc8_} \end{figure}

For the energy one cannot separate the singular and the analytic parts.
In addition, the singular part decays as $l^{\rho}$, with 
$\rho=\frac{ \alpha-1}{\nu}$. Using our estimate of 
$\frac{\alpha}{\nu}=-.536(32)$ and the hyperscaling relation 
$\frac{\alpha}{\nu}=\frac{2}{\nu}-d$, we find $\rho=-1.268(48)$.
Thus the variance of the singular part of the energy is expected according
to (\ref{eq: R scaling WD}) to scale as $l^{-3.072}$. Therefore one     
would expect $V_E$ to be dominated by the fluctuations
of the analytic part of $E$ decaying as $l^{-d}$.
In figure \ref{fig: VE_rc8_} the variance of the energy
$V_E$ as a function of $\log l$ is plotted.
Straight forward linear fits to the form $V_E= A_{vE} l^{x_E}$
in the lattice size range $24 \leq l \leq 256$ yielded $ x_E= -2.005(26)$
 in good agreement with our expectation $x_E=-d$.


To conclude this part of the study, we found weak self averaging at criticality for a disordered model governed by a pure fixed point with 
$\left(\frac{\alpha}{\nu}\right)_{\text{p}}<0$. We also found good agreement
 with the scaling prediction (\ref{eq: R scaling AH pure}).

 \section{The site dilute Ising model on a cubic lattice}
\label{sec: Ising}
The second model we chose to study is the site-dilute Ising
 model on a cubic lattice ( see e.g. \onlinecite{Sel Sing} and references 
 therein).
On every site of a $l \times l \times l$ cubic lattice either an Ising
 magnetic spin
$S_i=\pm 1$ is placed if $K_i=1$ or a vacancy is placed if $K_i=0$. The $K_i$ 
are randomly drawn according to one of the prescriptions given below.
 The system is governed by the Hamiltonian
\begin{equation}
{\cal H}= -J \sum_{<i,j>} K_i S_i K_j S_j \;,
\label{eq: Ham Is} \end{equation}
where $<i,j>$ stands for a pair of nearest neighbors. 
RG calculations found a dilution independent random fixed point with 
universal critical exponents. For  
example, a recent calculation\cite{Janssen 95} obtained $\gamma=1.313$, 
$\beta=0.342$ $\nu=0.666$ and $\alpha=0.002$.  Early MC studies found global 
effective  critical exponents which were found to depend on dilution.
This was later interpreted as due to crossover effects. For example,
in a most extensive MC study, Heuer\cite{Heuer 93} found from finite size 
scaling in the 
lattice size range $20\leq l \leq 60$ values ranging from 
$\frac{\alpha}{\nu}(p=0.95)=0.12(6)$  with 5\% vacancies, to 
$\frac{\alpha}{\nu}(p=0.8)=-0.04(6)$ and 
$\frac{\alpha}{\nu}(p=0.6)=-0.22(6)$. However he argued by analyzing a 
suitable scaling function, that all models with different amounts of 
dilution are exhibiting a crossover to the fixed point predicted by RG.
His results show that, of the amounts of dilution he studied, the $p=0.8$
model reached the universal behavior at the smallest lattice sizes.
Later Janssen Oerding and Sengespeick\cite{Janssen 95} showed in their RG 
calculations
that the effective exponent values obtained by Heuer can be related to
regions in the space of coupling coefficients away from the fixed point.

  
\subsection{Details of the simulations}
\label{sec: Ising simulation  details}
Three site-dilute Ising models were examined, including two types of disorder.
In one model disorder was realized in a canonical manner; namely, the
number of magnetic sites in each site-diluted sample was fixed at a 
 fraction $c=0.6$ of the number of sites in the lattice. Thus fluctuations
among samples occur only in the locations of the magnetic sites but not in 
their number. In two other models disorder was realized in a grand-canonical
 manner; namely, each sample was created by assigning to each site of the
lattice a magnetic spin (vacancy) with probability $p$ ($1-p$). In one model
we used $p=0.6$ and in the second one $p=0.8$. In this case fluctuations
among samples include fluctuations in the number of magnetic sites. 
These fluctuations tend to zero as $l \rightarrow \infty $, but for finite $l$
 they are significant. For this reason we found it of interest, in this
study of fluctuations among samples, to compare the two ensembles. We are 
unaware of any previous findings attesting to differences in the asymptotic 
critical behavior between the two ensembles. Because of the (spatially) 
uncorrelated nature of the disorder in the  grand canonical ensemble it is
favored for its relative simplicity by theoretical studies (see  
\onlinecite{Janssen 95} for references) and by numerical
studies\cite{Wang 89,Holey 90,Wang 90,Wyersberg 92,Hennecke 93a} aiming to 
test them. On the other hand, 
in studying by Monte Carlo {\em average} thermodynamic
observables, errors can be reduced by using canonical disorder, as was done 
in \onlinecite{Heuer 93}. 
We note that if one wishes to study by Monte Carlo
the {\em fluctuations} in the thermodynamic
observables due to disorder, the use of grand-canonical disorder is 
advantageous.

In the Monte Carlo simulations we used  the Wolff\cite{Wol:1C} single cluster
 algorithm\cite{Wang90a} for the Ising model because of its 
 efficiency\cite{Hennecke 93b}.
Skewed periodic boundary 
conditions\cite{BH:book} were used in order to speed up the simulations.
For each model and lattice size  $n$ site-dilute samples were simulated.
Table \ref{tab: n_samples} summarizes the number of samples $n$ used for each 
lattice size for the three models. 
Simulations were performed at the estimated infinite lattice critical
 temperatures $T_c^\infty$, given in table \ref{tab: n_samples}, due to
 Heuer\cite{Heuer 93}, and taking $J=1$.
The procedure for calculating the $T_c(i,l)$ is described in 
Sec. \ref{sec: calc Tc}.

 \begin {table}
\caption{  Number of site-dilute samples simulated for each model and lattice
 size $l$. The last column lists the infinite lattice critical temperature,
as estimated by Heuer\protect\cite{Heuer 93}, at which the simulations were
performed. For $c=0.6$, $l=90$ the pseudocritical temperature was not
estimated. }
 \label{tab: n_samples}
\begin{tabular}{llllllll}
              & $l=10 $  & $l=20$ & $l=40$ &$l=60$  &$l=80$&$l=90$   & $T_c^\infty$\\
\hline
 $c=0.6$      &  8000    & 26000  &  2000  &  800   & & 1000 & 2.4220(6)  \\  
 $p=0.6$      &  72000   & 47000  &  8000  &  950   & 800&   & 2.4220(6)  \\
\hline
\hline
              & $l=4$     & $l=8$ & $l=16$ & $l=32$ & $l=64$& &   \\
\hline
$p=0.8$       & 10000     & 4000  & 32000  & 4000 &  1479  & &  3.4992(5)  \\
 \end{tabular}
\end{table}
 \subsection{Finite size scaling  at $T_c^\infty$}

Here we give the finite size dependence of averages (over all samples)
of some observables at the critical point $[\overline{X}_i(T_c^\infty)]$.



\subsubsection{ Magnetization $m$ and susceptibility $\chi_c$ }
Using the abbreviation
\begin{equation}
M=\sum_i K_i S_i \;,
\label{eq: Def M} \end{equation}
the magnetization density $m$ is defined as
\begin{equation}
m=  \frac{[\av{|M|}]}{N p} \;,
\label{eq: m } \end{equation}
where $N=l^3$ and the fraction of magnetic sites was either $p=0.8$ or $p=0.6$.
The susceptibility at $T_c^\infty$ was defined as\cite{BH:book}
\begin{equation}
\chi_c=  \frac{[\av{M^2}]}{N p T} \;.
\label{eq: chi high} \end{equation}
The magnetization density $m$ and susceptibility $\chi_c$ were fitted to the
 finite size scaling forms (\ref{eq: m and Chi }).
\begin{figure}
\centering
\epsfxsize=87mm
\centerline{ \epsffile{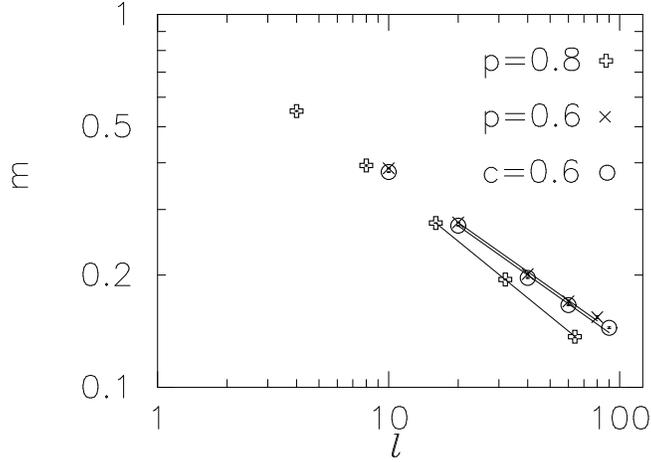} }
\caption{ The magnetization $m$ at $T_c^{\infty}$ as a function of 
$\log_{10}l$. The solid lines are fits to the form
 (\protect\ref{eq: m and Chi }), yielding estimates for
 $\frac{\beta}{\nu}$ which are listed in table \protect\ref{tab: Mag_exp}. 
 }
 \label{fig: M1_sum_} \end{figure}
\begin{figure}
\centering
\epsfxsize=87mm
\centerline{ \epsffile{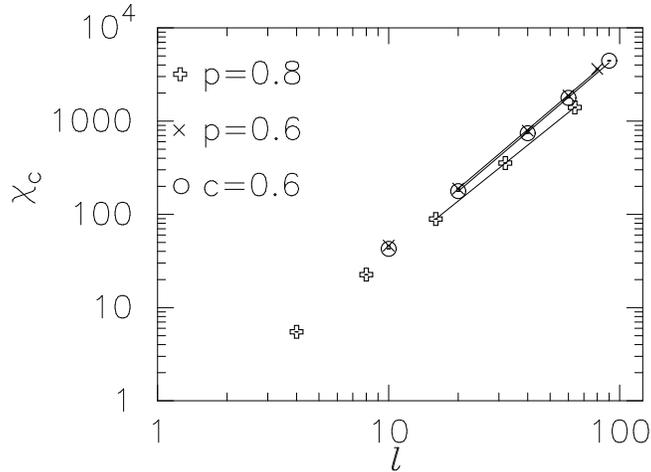} }
\caption{ The susceptibility $\chi_c$ as defined in
 (\protect\ref{eq: chi high}) at $T_c^{\infty}$ as a function of 
$\log_{10}l$. The solid lines are fits to the form
 (\protect\ref{eq: m and Chi }), yielding estimates for
 $\frac{\gamma}{\nu}$ which are listed in table \protect\ref{tab: Mag_exp}. 
 }
 \label{fig: Xt_sum_} \end{figure}

The estimates which were obtained for the critical exponents ratios 
$\frac{\beta}{\nu}$ and
$\frac{\gamma}{\nu}$ from the fits in figures \ref{fig: M1_sum_} and
\ref{fig: Xt_sum_} are listed in Table \ref{tab: Mag_exp}
together with the estimates of Heuer\cite{Heuer 93}. Note that exponent ratios
and critical temperatures quoted from \onlinecite{Heuer 93} for $p=0.8$ were
actually obtained with canonical disorder $c=0.8$.
 \begin {table}
\caption{  Estimates for order parameter critical exponent ratios from
 finite size scaling at $T_c^\infty$.   Estimates due to 
 Heuer\protect\cite{Heuer 93} are listed for comparison. Note that throughout
the paper the errors given for our results
are only statistical while Heuer's error estimates include the systematic
errors which could arise from errors in determining $T_c^\infty$. 
 }

\begin{tabular}{lllll}
      & $\frac{\beta}{\nu}$ &$\frac{\beta}{\nu}$  (Heuer)  &$\frac{\gamma}{\nu}$ &
                                                          $\frac{\gamma}{\nu}$(Heuer)\\
\hline
 $c=0.6$ & 0.438(13) & 0.45(2)& 2.110(21) & 2.09(3) \\  
 $p=0.6$ & 0.437(12)&        & 2.104(20) &         \\
 $p=0.8$ & 0.505(2) & 0.51(2)& 1.990(4) & 1.98(3) \\
 \end{tabular} \label{tab: Mag_exp} \end{table}


\subsubsection{ $\frac{\partial m}{\partial t}$ and estimation of 
$\frac{\alpha}{\nu}$ }
In attempting to estimate the exponent ratio $\frac{\alpha}{\nu}$ directly
from the finite size scaling of the specific heat $C$ through (\ref{eq: C}), we
 encountered two difficulties. First, we found that the estimates of the
 specific heat of each sample $\overline{C}_i$ were very sensitive to
the length of the simulations. Shorter simulations biased the specific heat
to lower values. e.g. for $p=0.6$, $l=40$ The value measured for $C$ using
average simulation length of $n/( 2\tau_E +1)\approx 120$ 
  was two standard deviations
smaller than the one measured with a four times longer simulation. The
 systematic underestimation of response functions due to run lengths which
 are
 too short was studied in \onlinecite{systematic}. Second, the accuracy in 
estimating $\frac{\alpha}{\nu}$ from the 
specific heat behavior is rather poor. This is due to the fact that 
$\frac{\alpha}{\nu}$ is a small negative number so that the singular
 behavior of $C$ is difficult to disentangle from other analytic
 contributions\cite{Heuer 90}.  

In order to overcome the difficulty in estimating the
exponent ratio $\frac{\alpha}{\nu}$ we followed Heuer\cite{Heuer 90} 
and measured the derivative of the magnetization with respect to the reduced
temperature $t=\frac{T-T_c}{T_c}$.  It is equal to the
magnetization-energy correlation
\begin{equation}
\Gamma=-\frac{\partial m}{\partial t}= - \frac{\partial^2 f}{\partial H 
             \partial t} \approx \frac{-1}{N p T^2} [ \av{ ( |M|-\av{|M|} )
                         ( {\cal H}- \av{{\cal H}} )} ] 
\;, \label{eq: gamma def} \end{equation}
where $f$ is the free energy density.
From the scaling behavior of the free energy 
\begin{equation}
f(t,h)= b^{-d} f(b^{y_t} t, b^{y_h} h )
\;\label{eq: f scaling} \end{equation} 
one finds that $\Gamma$ diverges as $t^{-\zeta}$, where 
\begin{equation}
\zeta=\frac{y_t-(d-y_h)}{y_t}= 1-\beta
\;.\label{eq: zeta} \end{equation}
Thus
we fit $\Gamma$ to the finite size scaling form
\begin{equation}
\Gamma= C_0 l^{\frac{\zeta}{\nu}}
\;.\label{eq: dmdt} \end{equation}

The resulting estimates for $ \frac{\zeta}{\nu}$ are given in table 
\ref{tab: P_exponents2}.
Assuming the hyperscaling relation $\frac{\alpha}{\nu}=\frac{2}{\nu}-d$
and using (\ref{eq: zeta}) the scaling relation
$\frac{\alpha}{\nu}=2(\frac{\zeta}{\nu}+\frac{\beta}{\nu})-d$ is obtained.
Using this relation, the results for $m$ and $\Gamma$ are utilized
in  table \ref{tab: P_exponents2} to give estimates for $\frac{\alpha}{\nu}$
which are much more accurate than those obtained from analysis of the
 specific heat results.


 \begin {table}
\caption{  Estimates for critical exponent ratios from finite size scaling at
 $T_c^\infty$.   Estimates due to 
 Heuer\protect\cite{Heuer 93} are listed for comparison. The estimate for 
$\frac{\alpha}{\nu}$ is based on the relation 
$\frac{\alpha}{\nu}=2(\frac{\zeta}{\nu}+\frac{\beta}{\nu})-d$.
 }
 \label{tab: P_exponents2}

\begin{tabular}{lllll}
   & $(\frac{\zeta}{\nu})$& $\frac{\zeta}{\nu}$ (Heuer) &$(\frac{\alpha}{\nu})$ & $\frac{\alpha}{\nu}$ (Heuer) \\
\hline
 $c=0.6$  & 0.948(6) & 0.94(2) &-0.228(28) &-0.22(6) \\  
 $p=0.6$  & 0.958(3) &         &-0.210(30) &         \\
 $p=0.8$  &0.962(4)  & 0.97(2) &-0.066(9)& -0.04(6)\\
 \end{tabular}
\end{table}

\section{Lack of self averaging at $T_c^\infty$}
\label{sec: self Ising}
In order to obtain the variance $V_X$ and the relative variance $R_X$
the same procedure and considerations as described in section 
\ref{sec: calc V_X} were used.

\begin{figure}
\centering
\epsfxsize=87mm
\centerline{\epsffile{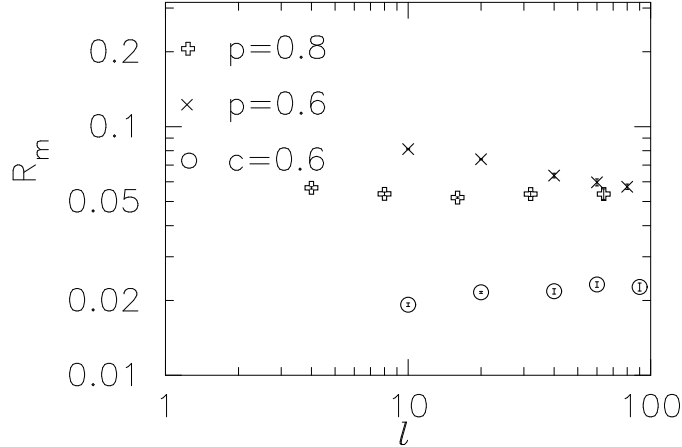} }
\caption{ The relative variance of the magnetization $R_m$ at 
$T_c^{\infty}$ as a function of $\log_{10}l$. 
 }
 \label{fig: RM_sum2} \end{figure}

In figure \ref{fig: RM_sum2} we plot the relative variance of the 
magnetization  $R_m$ as a 
function of lattice size on a double-logarithmic scale. 
Several interesting features are suggested by this figure. 
 First, note that for $p=0.6$, $R_m$  is decreasing as $l$ increases for the
 smaller lattice sizes, possibly leveling off for large $l$. $R_m$ of the
 $p=0.8$ model first decreases slightly and then seems to tend to a constant.
 Since
it seems plausible that $R_m(p=0.6) \geq R_m(p=0.8)$ for any lattice size,
 these trends seem to imply that for the two grand-canonical models
 $R_m$ tends
to the same constant. Assuming that this constant is bound from above by the
 $p=0.6$ model and from below by the $p=0.8$ model we estimate it
as $R_m=0.055(2)$. The implication of this scenario is that $R_m$
of the weakly diluted $p=0.8$ model reaches the universal $R_m$ value of the
 dilute Ising fixed
point  at smaller system sizes than the highly diluted $p=0.6$ model. The 
fluctuations in $m_i$ in the highly diluted $p=0.6$ model are larger than
those of the dilute Ising fixed point model. 
This finding is in line with Monte Carlo results\cite{Heuer 93} and
 RG calculations\cite{Janssen 95}, according
to which the critical exponents of the dilute Ising fixed point are
 closer, in the lattice size range
 $20 \leq l \leq 60$, to the observed effective critical exponents  of the $p=0.8$ model than to those of the $p=0.6$ model. 

A second feature is the striking
difference between the two types of disorder, with 
the canonically disordered $c=0.6$ model exhibiting a much smaller relative
 variance than that of the two  grand-canonical disorder models.
While $R_m$ of the $c=0.6$ model is initially increasing with system size it 
appears to level off to a constant value of $R_m(l=90)=0.0227(8)$. 
Though it is 
possible that $R_m$ could increase at larger lattice sizes it seems unlikely
since the system sizes are already quite large. An indicator to the similarity
of the two types of models $p=0.6$ and $c=0.6$ is the relative square root
mean of the fluctuations in the number of magnetic sites 
${\cal N}=\sum_{j=1}^N K_j$
in the $p=0.6$ systems
 $\sqrt{\frac{ [ ( {\cal N} - [{\cal N}])^2 ] }{ [{\cal N}]^2 } }= 
  \sqrt{ \frac{(1-p)}{N p  }}$ which for $l=80$ is as small as 
$\sim \frac{1}{1000}$.

 If indeed $R_m$ of the $c=0.6$ model tends to a different constant than
 that
of the models with grand canonical disorder, then according to Aharony and
 Harris' very general RG arguments\cite{Aharony 1996} the two types of models
 do {\em not} belong
to the same universality class! 
 We are not aware of any additional evidence
to this effect otherwise. For example our critical exponent estimates for
the $c=0.6$ and $p=0.6$ are compatible with each other, and our exponents
for the $p=0.8$ model are compatible with those of Heuer\cite{Heuer 93} for
 a $c=0.8$ model. The critical temperatures for both types of models
 seem also to agree (see table \ref{tab: K_c scaling} and References
 \onlinecite{Heuer 93,Wang 90}). 
This question is currently under investigation.
 Preliminary results\cite{AHW 1998} suggest that the two types of models
 do flow to the same
 fixed point and that the difference in $R_m$ will disappear for very large
$l$.

\begin{figure}
\centering
\epsfxsize=87mm
\centerline{ \epsffile{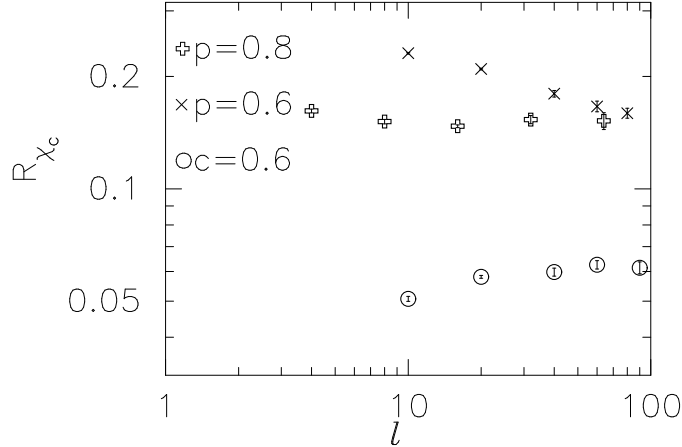} }
\caption{ The relative variance of the susceptibility $R_{\chi_c}$ at 
$T_c^{\infty}$ as a function of $\log_{10}l$. 
 }
 \label{fig: RXt_sum} \end{figure}

The relative variance of the susceptibility $R_{\chi_c}$ is plotted in Fig. 
\ref{fig: RXt_sum}. $R_{\chi_c}$ exhibits the same qualitative behavior as that
 of 
$R_m$. $R_{\chi_c}$ of the grand canonical disorder models tends to 
$R_{\chi_c}=0.156(4)$, while $R_{\chi_c}$ of the $c=0.6$ model seems to tend
 to  $R_{\chi_c}(l=90)=0.061(2)$. 
Aharony and Harris\cite{Aharony 1996,Aharony 1997} found that 
to leading order in $\epsilon=4-d$, 
$R_M/R_\chi= 1/4$. We find that for $p=0.8$
$R_M/R_\chi=0.35(2)$ and for $c=0.6$ $R_M/R_\chi=0.37(1)$. Possibly terms of 
higher order in $\epsilon$ would reconcile this discrepancy. 
It cannot be attributed to the definition
of $\chi_c$. If one defines the susceptibility as in (\ref{eq: chi low})
 then at $T_c^\infty$ one finds that $R_\chi$ becomes smaller by a factor of
 $\sim 7-10$. In this case the ratio $R_M/R_\chi$ would become even larger.
 We did not use this definition for the susceptibility at $T_c^\infty$ because
of its large single sample errors $\delta\overline{\chi}_i$ (see also
\onlinecite{BH:book}). 


\section{Scaling of pseudo critical temperatures}
\label{sec: Tc(i,l)}
\subsection{Calculating $T_c(i,l)$ with the histogram reweighting method}
\label{sec: calc Tc}
One of the main purposes of this work was to study the distribution
of pseudo-critical temperatures $T_c(i,l)$ of the ensemble of site-dilute
Ising models. The main aim was to study directly the scaling  of 
$\delta T_c(l)$ with $l$ and test which one of 
(\ref{eq: delta Tc}) or (\ref{eq: delta Tc AH}) is correct
in the case of a system governed by a disordered fixed point. The inverse
pseudo-critical temperature $K_c(i,l)=1/T_c(i,l)$ of the $i$'th sample was
 defined as 
 the inverse temperature of the maximum of the susceptibility of that sample,
$K_c(i,l)\equiv K_{\text{max}}(i,l)$. Here the definition of the 
susceptibility was
\begin{equation}
\chi_i=  \frac{\av{M^2}- \av{|M|}^2}{N p T} \;.
\label{eq: chi low} \end{equation}

In order to find $K_{\text{max}}(i,l)$ the following iteration procedure was followed
for each sample. A first simulation was performed at the infinite lattice
critical temperature (as estimated in \onlinecite{Heuer 93}) $K_1=K_c^\infty$
 (the index $i$ is omitted from here on).
In addition to calculating the observables $m, \chi, \Gamma$, a histogram of
the energy and magnetization was generated. Using the single histogram
reweighting technique\cite{Ferrenberg 88,Ferrenberg 91,Munger 91,Swendsen 93,Ferrenberg 95} (For previous studies of disordered systems utilizing the 
histogram reweighting technique see \onlinecite{Wyersberg 92,Chen 95}),
 this histogram can serve to calculate
 observables at temperatures close to $K_1$. By calculating $\chi$ at 
different temperatures a first estimate for the susceptibility maximum 
$\chi_{\text{max}}^1$ and the temperature at which it occurs 
$K_{\text{max}}^1$ was obtained. 
A second simulation was then performed at a temperature somewhat above
this estimate $K_2=K_{\text{max}}^1-offset$. Previous studies
of the histogram reweighting technique have shown that
 the errors of observables at $T$, $\delta\overline{X}(T)$, are 
smaller\cite{Ferrenberg 95,Munger 91}
when the temperature at which the histogram was generated $T_{\text{sim}}$ is
 slightly higher, $T_{\text{sim}}> T$. For this reason we chose $offset$ to be
 a small positive number (see more details below). 

Using the energy and magnetization histogram generated at $K_2$
a new estimate for the temperature of the susceptibility maximum, 
$K_{\text{max}}^2$, was obtained. If the difference between
the two estimates was smaller than a predetermined
resolution $r$, $|K_{\text{max}}^2 - K_{\text{max}}^1| < r$, the iteration process was
 stopped. Otherwise
 the iteration process continued, whereby $K_j=K_{\text{max}}^{j-1}-offset$, until the
 condition
\begin{equation}
  |K_{\text{max}}^j - K_{\text{max}}^{j-1}| < r 
\label{eq: converge}\end{equation}
 was met.  This iteration process was intended to overcome the problem of 
systematic errors\cite{Munger 91} that occur when the simulation temperature
 is too far from the true $K_{\text{max}}$. 
The condition (\ref{eq: converge}) is supposed to ensure that the last two 
estimates for $K_{\text{max}}$ do not suffer from a systematic error.
 $r$ was chosen equal to the approximately expected
statistical error of $K_{\text{max}}^j$. 
If the iteration process did not terminate
before or with the third estimate $K_{\text{max}}^3$ then the Monte Carlo
simulation length at $K_4$ was doubled and the process was continued. It
was again doubled if it reached the seventh iteration and again doubled
if it reached the tenth iteration. Non-convergence of the process 
after twelve iterations was very rare. In those samples the iteration procedure was restarted
manually with $K_1=K_c^\infty$ but with a larger initial Monte Carlo simulation length.
The need to increase the simulation length for some samples occured because
for different samples there were different auto-correlation times 
(of the Monte Carlo dynamics) and
different average cluster sizes, while the simulation length was specified by
the {\em number } of Wolff cluster flips.  

 In order to estimate the statistical error 
and reduce it, a simulation with five times as many  Monte Carlo steps
 (compared to the simulation length of the last iteration) was
 performed again at the last 
simulation temperature $K_j$. The Monte Carlo sequence was broken into five,
using each segment to create a separate histogram and calculate a separate
estimate of $K_{\text{max}}$ and $\chi_{\text{max}}$. Together with the last
estimates of $K_{\text{max}}$ and $\chi_{\text{max}}$ of the iteration
procedure, all together six estimates of $K_{\text{max}}$ and 
$\chi_{\text{max}}$ were averaged to give final estimates of $K_{\text{max},i}$
 and $\chi_{\text{max},i}$. The variance of these six estimates was used to estimate the error
for the two quantities, $\delta K_{\text{max},i}$, and 
$\delta \chi_{\text{max},i}$.

The parameter
$offset$ was adjusted  for the small system sizes, through trial runs,
 so as to minimize the errors in $\chi_{\text{max}}$, while its value
for the larger systems was extrapolated from the smaller ones. For $c=p=0.6$
 we set $offset\approx 0.27 l^{-1.66}$, and for $p=0.8$ $offset\approx 0.12 l^{-1.63}$. 
The optimal value of $offset$ was found not to 
depend strongly on the simulation length. 
The resolution $r$ was adjusted so as to be approximately equal to 
the ensemble average statistical error of $K_{\text{max}}^j$. 
Note  that the parameters $offset$ and $r$ were set once for each model
and each lattice size and were not varied for different samples.
In some of the larger systems, for a subset of the samples, the simulation with five times as many Monte Carlo steps was not performed, so that error estimates of $K_{\text{max}}$ and $\chi_{\text{max}}$ were not obtained. 
This was done in order to save computer time. For these samples the average
 squared error  of $K_{\text{max}}$ and $\chi_{\text{max}}$ was approximated
 as being six times larger 
than that of the complementary subset of samples where the error was calculated
(from an all together six times longer Monte Carlo sequence).
For the $p=0.6$ $l=80$ system the estimated average squared error was 
extrapolated from the smaller systems.
 

\subsection{Scaling of $t_c(l)$}
In the finite size scaling theory of \onlinecite{WD Self-Av} it was 
assumed that the average pseudo-critical temperature 
$K_{\text{max}}(l)\equiv [K_{\text{max}}(i,l)]$ scaled as
 \begin{equation}
K_{\text{max}}(l)-K_c = A_{K} l^{-\lambda}  \;,
\label{eq:  scale Kc}  \end{equation}
and that the shift exponent\cite{Barber} $\lambda=y_t=1/\nu$. 
 First we assumed the correctness
of the critical temperature values $T_c^\infty$ quoted in table 
\ref{tab: n_samples}, so
 that
the critical inverse temperature of the infinite sample  is assumed to be
$K_c=1/T_c^\infty=.285781(40)$ for $p=0.8$, and $K_c=0.41288(10)$ for $c=0.6$
 and $p=0.6$. 
Fitting $K_{\text{max}}(l)$ to (\ref{eq:  scale Kc}) with $K_c$ fixed
we found for the $p=0.8$ model values of $\lambda=1.446(34)$ and $A_{K}=0.040(5)$
from lattice sizes $16\leq l \leq 64$.
For the  $c=0.6$ and $p=0.6$ models the results were incompatible with the 
fixed value of $K_c=0.41288$. In fact in these models $K_{\text{max}}(l)$ monotonically decreases with $l$ and for the largest  lattices we have 
$K_{\text{max}}(80,p=0.6)-K_c=-0.00026(3)$ and 
$K_{\text{max}}(60,c=0.6)-K_c=-0.000077(35)$.
Thus 
we also fitted $K_{\text{max}}(l)$ to (\ref{eq:  scale Kc}) with $K_c$ being
a free parameter. The values of $\lambda$ and $K_c$ which were found,
using lattice sizes $10\leq l \leq 60$ for $c=0.6$ and $p=0.6$, and
$8 \leq l \leq 64$ for $p=0.8$, are given in
the third and fourth columns of table \ref{tab: K_c scaling}. 
For $p=0.8$ our estimate $K_c=0.2857609(4)$ is within errors of the estimate
of Heuer\cite{Heuer 93} (with canonical disorder) and of Wang et
 al\cite{Wang 90} (with grand canonical disorder). For $c=0.6$ and $p=0.6$
our estimates $K_c=0.41254(13)$ and $K_c=0.41251(5)$ are within errors
of each other but not within errors of the assumed value $K_c=0.41288(10)$.
A more accurate estimate of  $K_c$, which does not require knowledge of 
$\lambda$, is obtained in section \ref{sec: Tc estimate}.
In the fifth column of table \ref{tab: K_c scaling} estimates of $y_t$ based
 on estimates of $\frac{\beta}{\nu}$, $\frac{\zeta}{\nu}$ and 
the scaling relation (\ref{eq: zeta}) are given. The values of $y_t$ obtained
 by Heuer in the same way are given
in the sixth column of table \ref{tab: K_c scaling} and are compatible with
our estimates. 
Our estimates of $y_t$ and $\lambda$ agree for $p=0.8$ (where $K_c$ was fixed)
and for $p=0.6$ (where $K_c$ was a free parameter). For $c=0.6$ no agreement 
was found. One possible reason could be that the system size used to estimate
$\lambda$ was too small and that corrections to scaling need to be taken
into account. As is well known, finding the critical temperature
and the shift exponent simultaneously is a difficult task. In any case, our
 estimates for $y_t$ are much more accurate than the estimates for $\lambda$. 
It seems to us that trying to extract the shift exponent and the critical 
temperature by finding the 
pseudo-critical temperature of many samples and using their average 
$K_{\text{max}}(l)$ in 
(\ref{eq:  scale Kc}) is not an efficient method. This is because a long
MC simulation is needed to avoid systematic errors in estimating 
$K_{\text{max}}(i,l)$. Thus it is difficult to obtain a sufficient 
number of samples for an accurate enough estimate of $K_{\text{max}}(l)$.
The error in $K_{\text{max}}(l)$ must be small compared to 
$K_{\text{max}}(l)-K_c$.
This difficulty is not as significant for the estimate of the variance 
$V_{K_{\text{max}}}$.

 \begin {table}
\caption{  Different parameters related to the average pseudo-critical
inverse temperature $K_{\text{max}}(l)$ and its variance 
$V_{K_{\text{max}}}$. 
Second column: estimate of the shift exponent $\lambda$ according to
(\protect\ref{eq:  scale Kc}), where $K_c$ is taken from 
Heuer\protect\cite{Heuer 93}. Third and fourth column: same as first column
but with $K_c$ being a free parameter. Fifth column: estimate of $y_t$ based
on the finite size scaling of $m$ and $\Gamma$. Sixth column: same as fifth
according to \protect\cite{Heuer 93}. Last column: exponent of 
$V_{K_{\text{max}}}$.
 }
\begin{tabular}{lllllll}
   & $\lambda$, ($K_c$ fixed) & $\lambda$
   & $K_c$ &$y_t=\frac{\zeta}{\nu}+\frac{\beta}{\nu}$& $y_t$ (Heuer)&$\frac{\rho_K}{2}$ \\
\hline
 $c=0.6$& -----    & 0.99(19)& 0.41254(13)  &1.386(14)  & 1.39(4)& 1.41(4)\\  
 $p=0.6$& -----    & 1.30(10)& 0.41251(5)  & 1.395(12)  & ----   & 1.421(9) \\
 $p=0.8$& 1.446(34)& 1.346(2) & 0.2857609(4)& 1.467(5)& 1.47(4)& 1.44(2)  \\
 \end{tabular}
 \label{tab: K_c scaling} \end{table}
%


\subsection{Variance of pseudo-critical temperatures $V_{K_{\text{max}}}$}
\begin{figure}
\centering
\epsfxsize=87mm
\centerline{ \epsffile{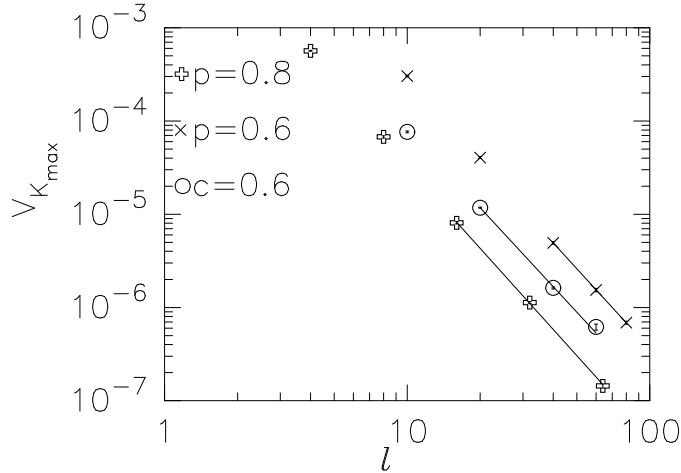} }
\caption{ The  variance of the inverse pseudo-critical temperatures 
$V_{K_{\text{max}}}$  
 as a function of $l$ on a $\log-\log$ scale. The lines are linear fits
yielding exponents $\rho_K$ listed in the last column of table 
\protect\ref{tab: K_c scaling}.  
 }
 \label{fig: VKM_sum} \end{figure}

The variance of the pseudo-critical temperatures distribution 
$V_{K_{\text{max}}}$ was calculated taking into account the errors $[(\delta K_{\text{max},i})^2]$. This is done in a manner completely analogous to
the discussion of $V_X$ in Sec. \ref{sec: calc V_X}. 
$V_{K_{\text{max}}}$ is plotted in Fig. \ref{fig: VKM_sum} on a double
 logarithmic scale. The solid lines are fits to the form 
 $V_{K_{\text{max}}}=A_K l^{-\rho_K}$ and the resulting estimates
 of  $\frac{\rho_K}{2}$ are listed  
in the last column of table \ref{tab: K_c scaling}. As one would expect, 
$V_{K_{\text{max}}}$ is smaller for $c=0.6$ than for $p=0.6$, and is the
 smallest for $p=0.8$.
We see that for all three models the results for $\rho_K$ exclude the
 possibility (\ref{eq: delta Tc}) that $\rho_K=d=3$. On the other hand 
$\frac{\rho_K}{2}$ is within
 errors of $y_t$ for $p=c=0.6$, and within errors of $\lambda$ (with $K_c$
fixed) for $p=0.8$,
as predicted by Aharony and Harris (\ref{eq: delta Tc AH}).
 Note that the 
values obtained for $\rho_K$ for $p=0.8$ with lattices $8 \leq l \leq 32 $
and $p=0.6$ with $20 \leq l \leq 60$ are $\rho_K=2.95(6)$ and $\rho_K=3.00(4)$.
This behavior of $V_{K_{\text{max}}}$ could be a manifestation of a crossover
 from pure (\ref{eq: delta Tc}) to dilute (\ref{eq: delta Tc AH}) critical
 behavior. On the other hand for the model with the canonical disorder
 the crossover is in the opposite direction since for $c=0.6$
with $10 \leq l \leq 40$,  $\rho_K=2.77(7)$.

\begin{figure}
\centering
\epsfxsize=88mm
\centerline{ \epsffile{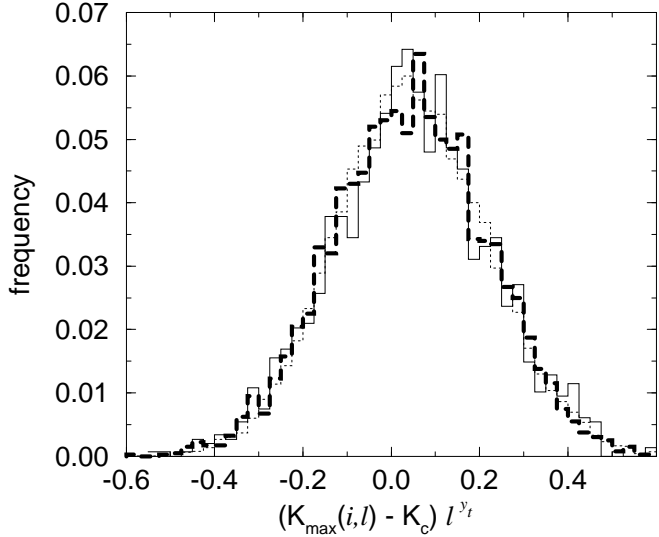} }
\caption{frequency distributions of the scaled pseudo-critical inverse 
temperatures $(K_{\text{max}}(i,L)-K_C) l^{y_t}$  for $p=0.8$, with 
$K_c=0.285781$ and $y_t=1.467$. Thin dotted line for $l=16$, thick dashed
 line for $l=32$ and thin solid line for $l=64$. The number of 
samples used was 32000 for $l=16$ 4000 for $l=32$ and 1479 for $l=64$.
 }
 \label{fig: jp8.histK} \end{figure}

 The results for $V_{K_{\text{max}}}$ support the picture implied by AH RG calculations, namely
that both the width of the pseudo-critical temperatures 
$\sqrt{V_{K_{\text{max}}}(l)}$ and the distance of its average from the 
critical inverse coupling $|K_{\text{max}}(l) -K_c|$ scale as $\sim l^{-y_t}$.
This is best visualized in Fig. \ref{fig: jp8.histK} where  the frequency of 
the scaled
 pseudo-critical inverse temperatures $(K_{\text{max}}(i,l)-K_C) l^{y_t}$ 
is plotted for $p=0.8$ and $l=16, 32, 64$ with $K_c=0.285781$
and $y_t=1.467$. It is evident that the three distributions match well.
Their averages are 
$[(K_{\text{max}}(i,l)-K_C) l^{y_t}]=0.047(1), 0.0451(28), 0.044(5)$,
and their widths are 
$\sqrt{V_{K_{\text{max}}}}\,l^{y_t}= 0.172(15), 0.174(28), 0.17(4)$ for
$l=16, 32, 64$ respectively. Note that the average ratio of the width to the
 average is $\approx3.8$ \ . Thus, as is evident from Fig. 
\ref{fig: jp8.histK}, the fluctuations in $K_c(i,l)$ are significantly larger
 than $|K_c(l)-K_c|$ for any system size. The result is that a measurement 
of $X$ at the critical temperature $T_c^\infty$ is done in some samples 
above their pseudo-critical temperature $T_c(i,l)$ and in some samples below
$T_c(i,l)$ ! 


\subsubsection{Estimating $T_c^\infty$ through $V_{K_{\text{max}}}$ }
\label{sec: Tc estimate}

\begin{figure}
\centering
\epsfxsize=88mm
\centerline{ \epsffile{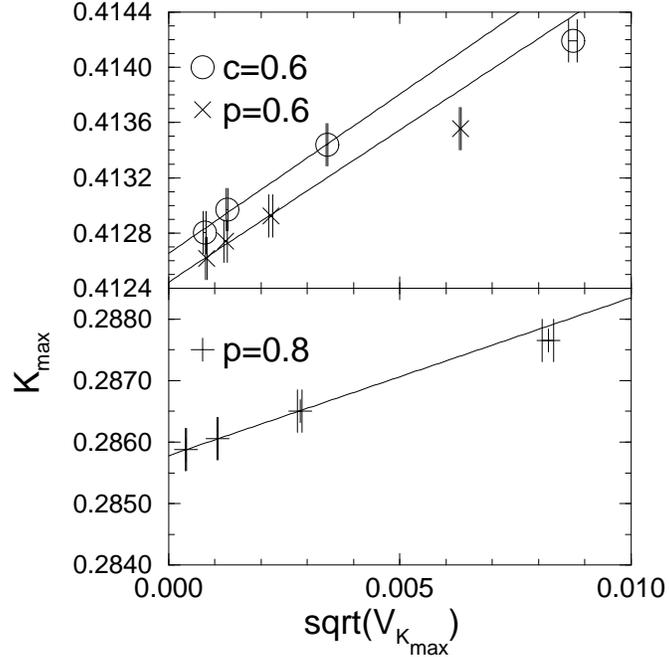} }
\caption{ The average pseudo-critical temperature $K_{\text{max}}$ as a 
function of the square root of $V_{ K_{ \text{max} } }$ together with linear
 fits made according to  (\protect\ref{eq: stauffer} ). Fits are made using
 the three largest 
system sizes for each model. The fitting parameters are listed in the text.
 }
 \label{fig: Kcboth} \end{figure}

Our estimates of $V_{K_{\text{max}}}$ allow us to estimate $T_c^\infty$ by
another method (We thank D. Stauffer for bringing this to our attention). 
Since asymptotically $K_{\text{max}}(l)- K_c \sim l^{-y_t}$
and $\sqrt{V_{K_{\text{max}}}} \sim l^{-y_t}$, one expects that
\begin{equation}
K_{\text{max}}(l)= K_c + B_v \sqrt{V_{K_{\text{max}}}(l)}
\;, \label{eq: stauffer}\end{equation}
where $K_c$ and $B_v$ need to be determined.
Note that by fitting the data according to (\ref{eq: stauffer}) (this method
was used in percolation studies\cite{Stauffer 1992}) it is not necessary to
 determine $\nu$, and 
only two fitting parameters are used. Therefore the estimates of $K_c$ 
obtained in this way are probably more reliable than those given in table 
\ref{tab: K_c scaling}. 
In figure \ref{fig: Kcboth} we plot $K_{\text{max}}$ as a function of
$\sqrt{V_{K_{\text{max}}}}$ together with linear fits made according to  
(\ref{eq: stauffer}). We find $K_c=0.285779(2)$, $B_v=0.257(3)$ for $p=0.8$,
$K_c=0.41244(4)$, $B_v=0.22(3)$ for $p=0.6$, and $K_c=0.41265(4)$, 
$B_v=0.230(15)$ for $c=0.6$.

\subsection{Maximum of the susceptibility $\chi_{\text{max}}$}

\subsubsection{Scaling of $[\chi_{\text{max}}]$ }
\begin{figure}
\centering
\epsfxsize=88mm
\centerline{ \epsffile{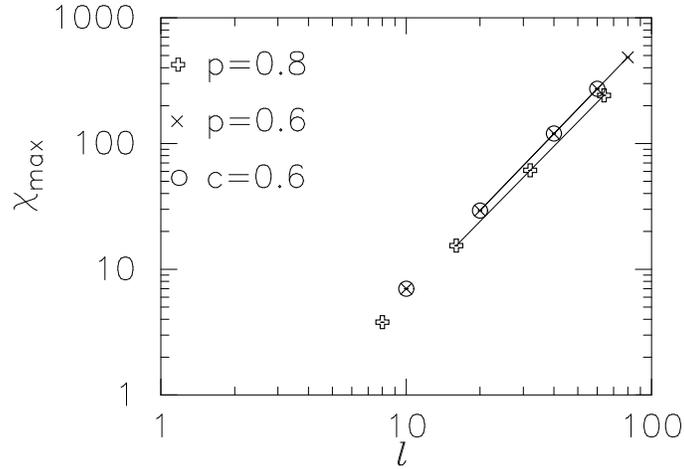} }
\caption{ The ensemble average of the maximum susceptibility
 $[\chi_{\text{max}}]$  as a function of 
$\log_{10}l$. The solid lines are fits to the form
 (\protect\ref{eq: m and Chi }), yielding estimates of 
$\frac{\gamma}{\nu}= 2.034(2)$ for $c=0.6$, $\frac{\gamma}{\nu}= 2.027(2)$
 for $p=0.6$, and  $\frac{\gamma}{\nu}= 1.987(3)$ for $p=0.8$. 
 }
 \label{fig: XM_sum_} \end{figure}

Another way to study the finite size scaling of the susceptibility is
to study the ensemble average of the maximum susceptibility
 $[\chi_{\text{max}}]$ which is expected to scale with lattice size as in
(\ref{eq: m and Chi }) with a scaling exponent $\frac{\gamma}{\nu}$.
In figure \ref{fig: XM_sum_} $[\chi_{\text{max}}]$ is plotted as a function
of $l$ on a double logarithmic scale. The straight lines are linear fits
to the form (\ref{eq: m and Chi }). For  $p=0.8$ we find 
$\frac{\gamma}{\nu}= 1.987(3)$ which is in agreement with
 the estimate obtained using the susceptibility $\chi_c$ at $T_c^\infty$,
 $\frac{\gamma}{\nu}=1.990(4)$. As can be seen in Fig. \ref{fig: XM_sum_}
the values of $[\chi_{\text{max}}]$ for $p=0.6$ and $c=0.6$ are 
indistinguishable (they are indeed within errors). This is in contrast with 
the data at $T_c^\infty$
of fig. \ref{fig: Xt_sum_} where $\chi_c$ of the two models seem to diverge
 with a similar exponent but with a different amplitude. We have also 
calculated $\chi$ at $T_c^\infty$ and found the same trend, namely that 
$\chi(p=0.6) > \chi(c=0.6)$, so that this feature is not an artifact
 of the different definitions, (\ref{eq: chi high}) and (\ref{eq: chi low}),
 for $\chi$. 
For $p=0.6$ and  $c=0.6$ we found $\frac{\gamma}{\nu}= 2.027(2)$
and $\frac{\gamma}{\nu}= 2.034(2)$ respectively. These values are 
significantly lower than the values found using the susceptibility
 $\chi_c$ at $T_c^\infty$, $\frac{\gamma}{\nu}=2.104(20)$ and
 $\frac{\gamma}{\nu}=2.110(21)$. They are also closer to results of
RG calculations $\frac{\gamma}{\nu}=1.97$\cite{Janssen 95,Mayer 89}. 

\subsubsection{ Lack of self averaging of the relative variance
 $R_{\chi_{\text{max}}}$}
\begin{figure}
\centering
\epsfxsize=88mm
\centerline{ \epsffile{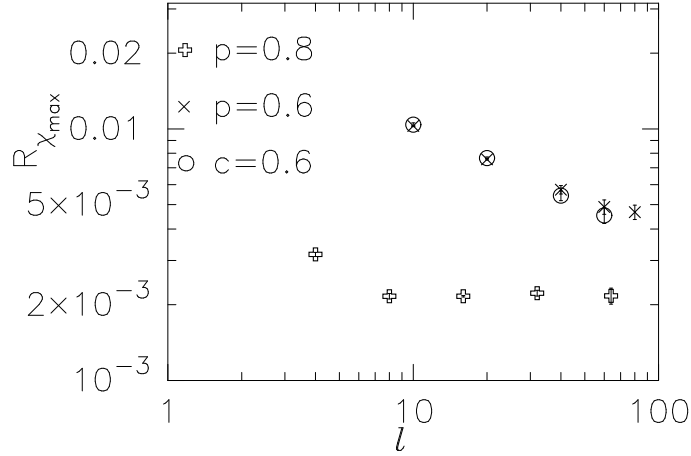} }
\caption{ The relative variance of the maximum susceptibility
 $R_{\chi_{\text{max}}}$ as a function of $\log_{10}l$. 
 }
 \label{fig: RXM_sum} \end{figure}

In figure \ref{fig: RXM_sum} We plot the relative variance of the 
maximal susceptibility  $R_{\chi_{\text{max}}}$ as a 
function of lattice size on a double-logarithmic scale.
For $p=0.8$, $l=64$ and $p=0.6$, $l=80$ we have
$[ ( \delta \overline{\chi_{\text{max},i}})^2 ]/V_{\chi_{\text{max}}} \approx \frac{1}{2}\,,\, 0.15$ respectively (see section \ref{sec: calc V_X}). 
Thus the estimate of
$V_{\chi_{\text{max}}}$ is dominated by the estimate of the average squared 
single sample errors $[ ( \delta \overline{\chi_{\text{max},i}})^2 ]$.
 For $p=0.6$ $l=80$ this estimate was actually 
extrapolated from the smaller systems estimates (see \ref{sec: calc Tc}). Thus
these two data points should be taken with more than a grain of salt.
It is most interesting to compare Fig. \ref{fig: RXM_sum} to Fig.
 \ref{fig: RXt_sum} where the relative variance
of the susceptibility at $T_c^\infty$, $R_{\chi_c}$, is plotted. For $p=0.8$ the behavior
of $R_{\chi_{\text{max}}}$ and $R_{\chi_c}$ is qualitatively rather similar
with $R_{\chi_{\text{max}}}$ initially decreasing as $l$ increases and tending
for larger $l$ to a constant, where  $R_{\chi_{\text{max}}}(l=64) =.00216(16)$.
However,  in contrast with Fig. \ref{fig: RXt_sum}, this constant is roughly
72 times smaller than the large $l$ value of $R_{\chi_c}$.
This is quite a striking difference. 
It means that in order to obtain the same
 relative accuracy in $[\chi_c]$ as in $[\chi_{\text{max}}]$ approximately 70 
times as many samples are needed.
The source of this difference is apparently simple.
 The susceptibility of
each sample is some function $G$ of the temperature with a sharply peaked
maximum at $T_c(i,l)$. In fact $G$ is approximately only a function of the difference $T-T_c(i,l)$, $G(T-T_c(i,l))$.
Thus, the value of the maximum susceptibility
 is nearly sample independent, $\chi_{\text{max}}\approx G(0)$. On the other
hand, when one measures $\chi$ at
$T_c^\infty$, in different samples one is sampling $G$ at different values of its argument. This results in large fluctuations in $\chi$ at $T_c^\infty$.

Our findings
 suggest that the standard procedure of investing much computation time
 in finding the $l \rightarrow \infty$ limit of the critical temperature,
$T_c^\infty$, and then averaging quantities at this temperature over many 
samples, is not optimal. A better procedure may be to locate
through the single or multiple histogram method the pseudocritical temperature
of each sample, and measure quantities at that temperature. In this way sample
to sample fluctuations are reduced substantially and better accuracy is 
achieved.

For $p=0.6$  $R_{\chi_{\text{max}}}$ monotonically decreases with lattice size,
possibly leveling off to a constant for large $l$. In contrast with
$R_{\chi_c}$, this constant is different from that of the $p=0.8$ model. 
Lastly, for $c=0.6$, in contrast with $R_{\chi_c}$, we find that 
$R_{\chi_{\text{max}}}$ is within errors of $R_{\chi_{\text{max}}}$ of the
$p=0.6$ model. In addition, $R_{\chi_{\text{max}}}$ initially decreases as $l$
 increases, opposite to the behavior of $R_{\chi_c}$. 
 More explanations to the differences in the behavior of 
$R_{\chi_{\text{max}}}$ and $R_{\chi_c}$ are given at the end of the
next subsection.


\subsection{ Dependence of $m(T_c^\infty)$ on $T_c(i,l)$ }

\begin{figure}
\centering
\epsfysize=4.25truein
\epsfxsize=5.truein
\input{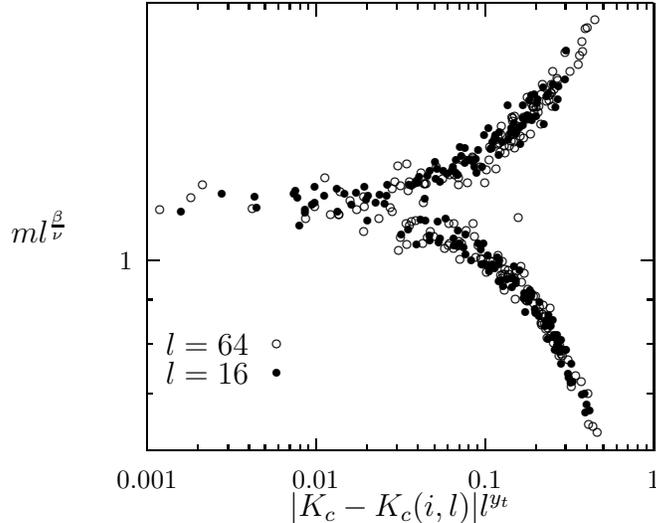}
\caption{ Scatter plot, where for each sample $i$ the horizontal axis 
represents the scaled absolute 
 inverse temperature $Z=|K_c - K_c(i,l)| l^{y_t}$ and the vertical axis is the 
scaled magnetization $\tilde{Q}_\pm= m l^{\frac{\beta}{\nu}}$. Points with 
$K_c> K_c(i,l)$ 
constitute the higher $m$ (lower temperature) branch, whereas points with 
$K_c< K_c(i,l)$ constitute the lower $m$ (higher temperature) branch.
 For
the sake of clarity, only  100 points are shown for each system size and each branch,
 and  several points with $|K_c - K_c(i,l)| l^{y_t} < 0.001$ were 
omitted. 
 }
 \label{fig: jp8.M.16_64.data} \end{figure}

\begin{figure}
\centering
\epsfysize=4.25truein
\epsfxsize=5.truein
\input{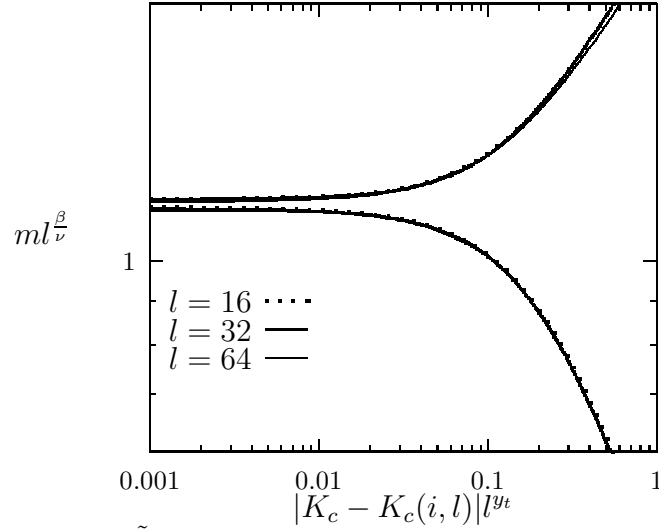}
\caption{ The functions $\tilde{Q}_\pm(Z)$, as defined in 
(\protect\ref{eq: Q form}), obtained from best fits to the scaled 
magnetization versus temperature scatter plots for $l=16,32,64$. upper curves
 according to $\tilde{Q}_-$ ($K_c>K_c(i,l)$) and lower curves according to 
$\tilde{Q}_+$ ($K_c< K_c(i,l)$). The fitting parameters are given in
table \protect\ref{tab: Qparam}. 
 }
 \label{fig: jp8.M.3fits} \end{figure}

After examining the behavior of the distribution of $X(i,l)$ at $T_c^\infty$
and the distribution of $T_c(i,l)$ it is imperative to examine the correlation
between the two distributions. A good starting point is the finite size 
scaling ansatz (\ref{eq: disorder fss}), according to which $X_i(T_c^\infty)$
mainly depends on $\frac{T_c^\infty-T_c(i,l)}{T_c^\infty}$. Fig. 
\ref{fig: jp8.M.16_64.data} is a scatter
plot where for each sample $i$ the horizontal axis represents the scaled absolute 
 inverse temperature $|K_c - K_c(i,l)| l^{y_t}$ and the vertical axis is the 
scaled magnetization $m_i l^{\frac{\beta}{\nu}}$. This representation is 
equivalent to the usual data collapse representation which is used to 
demonstrate finite size scaling. 
The difference is that here the reference critical temperature is $K_c(i,l)$ instead of 
$K_c$, and the measurement temperature is always $K_c$ instead of different values of 
$K$.
Points with $K_c> K_c(i,l)$ constitute the 
higher $m$ (lower temperature) branch, whereas points with $K_c< K_c(i,l)$ 
constitute the lower $m$ (higher temperature) branch. In figure 
\ref{fig: jp8.M.16_64.data} we plot data for $p=0.8$ and $l=16, 64$. For
the sake of clarity, only  100 points are shown for each system size and each branch,
 and  several points with $|K_c - K_c(i,l)| l^{y_t} < 0.001$ were 
omitted.  
Figure \ref{fig: jp8.M.16_64.data} indicates that to a good 
approximation the scaled magnetization of the sample at $T_c^\infty$ 
is a function of only the scaled reduced temperature of the sample. Thus 
one may attempt to substitute (\ref{eq: disorder fss}) by a sample independent
form for $\tilde{Q}_i(Z)$ so that
\begin{equation}
X_i(T,l)\approx l^\rho \tilde{Q}( \dot{t}_i l^{y_t} )
\;. \label{eq: disorder fss_noi} \end{equation}
Note that this is only a good {\em approximation}; if 
(\ref{eq: disorder fss_noi}) were exact, it would mean that 
$R_{\chi_{\text{max}}}=0$.
Thus in order to describe the magnetization data at $K_c$ we write
 (the change from temperature to inverse temperature is only for 
convenience)
\begin{equation}
m_i(K,l)= l^{-\frac{\beta}{\nu}} \tilde{Q}_{\pm}\{ |(K - K_c(i,l))| l^{y_t} \}
\;. \label{eq: m fss} \end{equation}
Here $\tilde{Q}_{+}(Z)$ is the scaling function for $K < K_c(i,l)$ and 
$\tilde{Q}_{-}(Z)$ for 
$K > K_c(i,l)$.
For large $l$, and thus large $Z$, the infinite sample critical behavior,
$m_i\sim \left\{K - K_c(i,l)\right\}^{\beta}$,
 must be asymptotically reproduced\cite{Landau 76} for $K > K_c(i,l)$. Thus,
 for large $Z$, $\tilde{Q}_{-}(Z)\sim Z^{\beta}$. 
For $K < K_c(i,l)$ the  shape of $\tilde{Q}_{+}(Z)$ must reproduce, for
large $Z$,
the $\frac{1}{\sqrt{N}}$ decay of the magnetization to zero as 
$l\rightarrow \infty$. Thus, for large $Z$,
 $\tilde{Q}_{+}(Z)\sim Z^{\beta-\frac{d}{2 y_t}}$. 
For $K=K_c(i,l)$, i.e. $Z=0$, the finite size scaling behavior 
$[m_i]\sim l^{-\frac{\beta}{\nu}}$ must be asymptotically 
reproduced\footnote{An example of this
type of finite size scaling is the scaling we found for 
$[\chi_{\text{max}}]$.}, implying 
 $\tilde{Q}_\pm (Z) \rightarrow\mbox{const}$ as $Z\rightarrow 0$. A simple possible form for 
$\tilde{Q}_\pm (Z)$
fulfilling these requirements is 
\begin{equation}
\tilde{Q}_\pm (Z)= A_\pm Z^{\rho_\pm} 
( 1 + B_\pm Z^{-p_\pm} )^{\frac{\rho_\pm}{p_\pm}}
\;, \label{eq: Q form} \end{equation} 
 where $\rho_-=\beta$ and $\rho_+=\beta-\frac{d}{2 y_t}$ and 
$A_{\pm}, B_{\pm}, p_{\pm}$ are free parameters, so that
 the data of Fig. \ref{fig: jp8.M.16_64.data} should be 
 described by
(\ref{eq: m fss},\ref{eq: Q form}) with $K=K_c$. Thus, for each lattice size 
$l=16,32,64$
and both branches, $K_c < K_c(i,l)$ and $K_c > K_c(i,l)$,  the scaled 
$\{K_{\text{max}}(i,l),m_i(K_c)\}$ pairs (a partial set of which is plotted 
in figure \ref{fig: jp8.M.16_64.data}) were fitted to the form 
(\ref{eq: Q form}). The values of $\rho_-=\beta=0.34295$ and 
$\rho_+=\beta-\frac{d}{2 y_t}=-0.67565$ which were used rely on
the finite size scaling results at $T_c^\infty$ (tables \ref{tab: Mag_exp}
 and  \ref{tab: K_c scaling}). The six fitting functions which were 
obtained are plotted in Fig. \ref{fig: jp8.M.3fits} and their fitting 
parameters are given in Table \ref{tab: Qparam}. 
 The agreement between the three curves for both branches, as seen in Fig. 
\ref{fig: jp8.M.3fits}, is surprisingly good. The goodness of the fits
is also extremely high. This suggests that equation 
(\ref{eq: disorder fss_noi})   , equations 
similar to (\ref{eq: Q form}), and the possibly invariant 
(as suggested by Fig. \ref{fig: jp8.histK} )
distributions of $K_c(i,l)$ provide an excellent description of the scaling
behavior of disordered systems.

 \begin {table}
\caption{ parameters of the fitting functions $\tilde{Q}_\pm (Z)$, defined in  
(\protect\ref{eq: Q form}), obtained  
by fitting the data sets of scaled $\{K_{\text{max}}(i,l),m_i(K_c)\}$ pairs 
for each lattice size $l=16,32,64$ separately (with p=0.8). The number of 
samples used
was 32000, 4000, and 1477 for $l=16,32,64$ respectively. 
The six fitting functions are plotted in Fig. \protect\ref{fig: jp8.M.3fits}.
The crossover lengths, which control the crossover to the large $Z$ behavior
are defined as 
$Z_\pm^{\mbox{cross}}=B_\pm^{\frac{1}{p_\pm}}$.
} 
\begin{tabular}{lllllllll}
   & $A_-$ & $B_-$ & $p_-$ &$Z_-^{\mbox{cross}}$& $A_+$ & $B_+$ & $p_+$ 
&$Z_+^{\mbox{cross}}$\\
\hline
$l=16$& 2.387(4) & 0.0518(13)& 1.45(1)  & 0.130& 0.4695(12) & 0.1765(18)&1.299(5)& 0.263 \\  
$l=32$& 2.380(9) & 0.048(3)  & 1.49(3)  & 0.130& 0.4623(22) & 0.1687(34)&1.316(9)& 0.259\\   
$l=64$& 2.291(23)& 0.07(1)   & 1.33(6)  & 0.142& 0.4524(75) & 0.152(11) &1.368(35)&0.252\\
 \end{tabular}
  \label{tab: Qparam} \end{table}

\begin{figure}
\centering
\epsfysize=4.25truein
\epsfxsize=5.truein                                           
\input{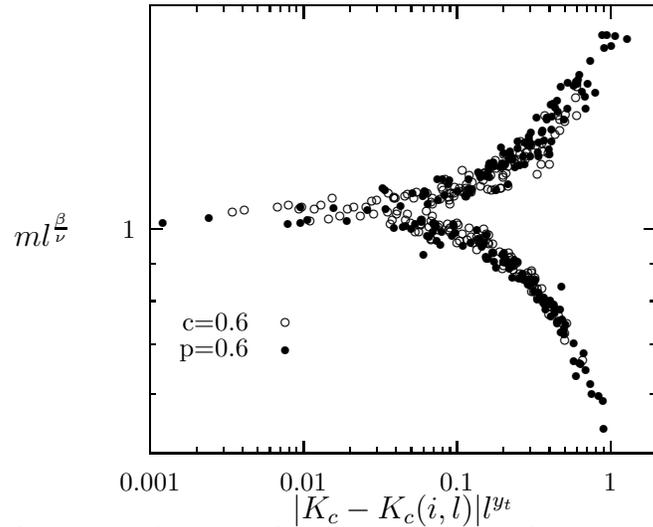}
\caption{ Same data as in figure \protect\ref{fig: jp8.M.16_64.data}  but for
 $p=0.6$ and $c=0.6$ with system size
$l=40$. The data were scaled with exponents taken as the average of the 
exponents of the two models, $\frac{\beta}{\nu}=0.4375$ and $y_t=1.3905$.
 For the sake of clarity, only a 100 points for each model and each branch
 are shown. 
}
 \label{fig: pc6.M.40.data} \end{figure}

In figure \ref{fig: pc6.M.40.data} we show the same data as in figure 
\ref{fig: jp8.M.16_64.data}  
but for $p=0.6$ and $c=0.6$ with system size
$l=40$. 
The purpose of this analysis is to demonstrate that the magnetization of the two 
models is governed by the same temperature dependence, and that the main
difference is in the distributions of $K_c(i,l)$. For this reason the data
 were scaled with the same exponents, taken as the average of the 
exponents of the two models, $\frac{\beta}{\nu}=0.4375$ and $y_t=1.3905$.
In fact our estimates for $y_t$ and $\frac{\beta}{\nu}$ for the two models are 
within errors.
 For the sake of clarity, only  100 points for each model and each branch
 are shown.  As was seen with the $p=0.8$ data, it is evident that to a good 
approximation in both models
the magnetization at $K_c$ is a function of only the reduced
inverse temperature  $K_c-K_c(i,l)$. The main difference between the two
models is also clear; For $p=0.6$ there are more points with large
$|K_c-K_c(i,l)|$, while for $c=0.6$ there are more points with small
$|K_c-K_c(i,l)|$. 
Thus larger fluctuations for $p=0.6$ in $K_c(i,l)$ (see also Fig. 
\ref{fig: VKM_sum}) together with the large dependence of $m_i(K_c)$
on $K_c-K_c(i,l)$ give rise to the result that $R_m(p=0.6)> R_m(c=0.6)$.

\begin{figure}
\centering
\epsfysize=3.25truein
\epsfxsize=4.truein
\epsfxsize=88mm
\centerline{ \epsffile{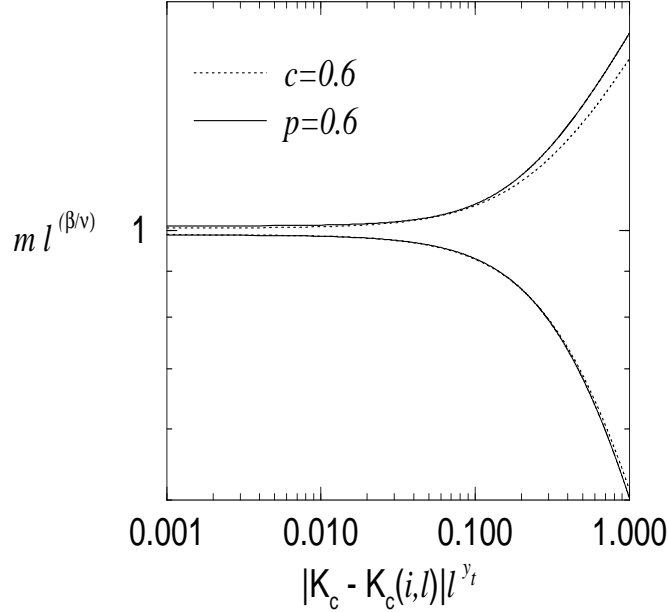} }
\caption{ The functions $\tilde{Q}_\pm(Z)$, as defined in 
(\protect\ref{eq: Q form}), obtained from best fits to the scaled 
magnetization verses temperature scatter plots for $c=0.6$ (dotted line) and
$p=0.6$, with $l=40$. upper curves
 according to $\tilde{Q}_-$ ($K_c>K_c(i,l)$) and lower curves according to 
$\tilde{Q}_+$ ($K_c< K_c(i,l)$). Fits made using $\frac{\beta}{\nu}=0.4375$
 and $y_t=1.3905$. 
 }
 \label{fig: jpc6.M.Scale.l40} \end{figure}

In figure \ref{fig: jpc6.M.Scale.l40} we plot the fitting functions 
$\tilde{Q}_\pm(Z)$, obtained by best fits to the scaled magnetization verses
 temperature scatter plots for $p=0.6$ and $c=0.6$ with
$l=40$ (the full data sets corresponding to figure \ref{fig: pc6.M.40.data}).
For the high temperature branch ($K_c< K_c(i,l)$, lower curve) good agreement
 between the fitting functions $\tilde{Q}_+(Z)$ of the two models is found. 
For the low temperature branch ($K_c> K_c(i,l)$, higher curve) good agreement
 is found between the functions $\tilde{Q}_\pm(Z)$ for smaller $Z$, while for
 large $Z$ $\tilde{Q}_-(Z)$ is larger for the grand canonical disorder 
(p=0.6). The fitting functions $\tilde{Q}_\pm(Z)$ for the data for $l=60$ did
 not agree with those of $l=40$.   
 Possibly this is so because the exponents used are not the asymptotic 
ones\onlinecite{Heuer 93,Janssen 95}. 



\begin{figure}
\centering
\epsfysize=3.25truein
\epsfxsize=5.truein
\epsfxsize=88mm
\centerline{ \epsffile{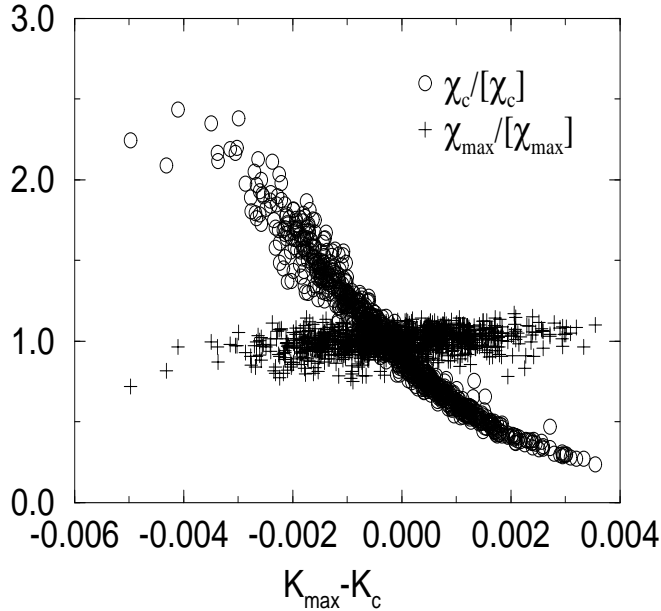} }
\caption{ A scatter plot of
 $(K_{\text{max}}-K_c,\frac{\chi_c(K_c)}{[\chi_c(K_c)]})$ and
 $(K_{\text{max}}-K_c,\frac{\chi_{\text{max}}}{[\chi_{\text{max}}]})$,
contrasting the dependence of $\chi_c(K_c)$
and $\chi_{\text{max}}$ on $K_{\text{max}}(i,l)$. Data for $p=0.6$
and system size $l=60$ from 950 samples. 
 }
 \label{fig: jq6.l60.XXM_vsKm} \end{figure}

It is also of interest to contrast the dependence of $\chi_{\text{max}}$
on $K_{\text{max},i}$ with the dependence of $\chi_c(K_c)$  on  
$K_{\text{max},i}$.
This is a key to understanding the reasons for the differences between
the  characteristics of $R_{\chi_c}$ (figure \ref{fig: RXt_sum}) and 
the characteristics of $R_{\chi_{\text{max}}}$ (figure \ref{fig: RXM_sum}).
In figure \ref{fig: jq6.l60.XXM_vsKm} we show a scatter plot of
 $(K_{\text{max}}-K_c, \frac{\chi_{\text{max}}}{[\chi_{\text{max}}]})$  and
 $(K_{\text{max}}-K_c, \frac{\chi_c(K_c)}{[\chi_c(K_c)]})$ for $p=0.6$
and system size $l=60$ from 950 samples. 
It is evident that while $\chi_c(K_c)$ shows a strong dependence on
$K_{\text{max}}-K_c$, $\chi_{\text{max}}$ shows little dependence on 
$K_{\text{max}}-K_c$. This qualitative difference persists for all models and
 all system sizes. This explains why, for any given model, fluctuations in 
 $K_{\text{max},i}$ give rise to fluctuations in $\chi_c(K_c)$ which are much
 larger than the fluctuations in $\chi_{\text{max}}$. The result is that
$R_{\chi_{\text{max}}}\ll R_{\chi_c}$, as we have noted previously.
 
Fig. \ref{fig: jq6.l60.XXM_vsKm} is also the key to understanding why  
$R_{\chi_c}(p=0.6)>R_{\chi_c}(c=0.6)$ while 
$R_{\chi_{\text{max}}}(p=0.6) \approx R_{\chi_{\text{max}}}(c=0.6)$.
In the first case, since fluctuations in $K_{\text{max}}$ are larger for
 $p=0.6$
than for $c=0.6$ (see figure \ref{fig: VKM_sum}) the strong dependence of
$\chi_c(K_c)$ on $K_{\text{max}}-K_c$ gives rise to 
$R_{\chi_c}(p=0.6)>R_{\chi_c}(c=0.6)$. In the second case, despite the fact
 that fluctuations in $K_{\text{max}}$ are larger for $p=0.6$
than for $c=0.6$, the weak dependence of
$\chi_{\text{max}}$ on $K_{\text{max}}-K_c$ results in 
$R_{\chi_{\text{max}}}(p=0.6) \approx R_{\chi_{\text{max}}}(c=0.6)$.

\section{Summary and discussion}
\label{sec: sum}
By and large it seems that our MC results confirm the AH scenario. 
In an Ashkin-Teller model, governed by a pure fixed point, we found
that $R_X\sim l^{\frac{\alpha}{\nu}}$. in agreement with
(\ref{eq: R scaling WD}) and (\ref{eq: R scaling AH pure}).
In site dilute Ising models on a cubic lattice, governed by a random fixed point, we found a lack of self averaging
for both  canonical and grand canonical disorder. 
One of the aims of our work was to resolve whether at random fixed points
our assumption (\ref{eq: delta Tc}), which led to the prediction
(\ref{eq: R scaling WD}) for the critical width $R_X$, is correct? The
alternative $R_X \rightarrow \mbox{const}$ result of AH implies that 
(\ref{eq: delta Tc AH}) should replace (\ref{eq: delta Tc}). Our results
indicate that the AH result is the correct one.
Note though that the absolute
value of the exponent ratio $\frac{\alpha}{\nu}$ of the dilute Ising fixed
point, either as calculated by RG, $\frac{\alpha}{\nu}=0.003$, or as indicated by the
 $p=0.8$ results $\frac{\alpha}{\nu}=-0.055(8)$, is very 
small. Thus one could argue that our results for
$R_m$ and $R_{\chi_c}$ do not disprove (\ref{eq: R scaling WD}). 
The scaling of $V_{K_{\text{max}}}$ is, however, in agreement with 
(\ref{eq: delta Tc AH})
and not with (\ref{eq: delta Tc}). This therefore rules out 
(\ref{eq: R scaling WD}) since it is based on (\ref{eq: delta Tc}).

We find it appropriate to repeat here the results of Aharony and 
Harris\cite{Aharony 1996}, which we have now validated,
with an emphasis on the implication to experiments.
 In finite size scaling form the relative variance can be written as
\begin{equation}
R_X(\xi,l)=l^\omega Q(l/\xi) 
\;. \label{eq: R sum}\end{equation}
For a fixed $\xi=\xi_0$ and $l>>\xi$, and thus large $Z$, strong
 self averaging, $R_X(\xi_0,l)\sim l^{-d}$, must be asymptotically reproduced.
 Thus
 $Q(Z)\sim Z^{-d-\omega}$ for large $Z$. At criticality the correlation length diverges and
\begin{equation}
\lim_{\xi \rightarrow \infty}R_X(\xi,l)=l^\omega Q(0) 
\;. \label{eq: R fss}\end{equation}
When the system is governed by a disordered fixed point 
$\omega=0$. When the system is governed by a pure fixed point 
$\omega=\left(\frac{\alpha}{\nu}\right)_p$.
Thus the two possible behaviors for $1\ll\xi\ll l$ are
   \begin{equation}
R_X(\xi,l)\sim \left\{ \begin{array}{ll}
\, \left(\frac{l}{\xi}\right)^{-d} & \mbox{for a random fixed point }  \\
\, \left(\frac{l}{\xi}\right)^{-d} \xi^{\frac{\alpha}{\nu}}& \mbox{for a pure fixed point }
  \end{array}
\right. \label{eq: R(xi)} \;.   \end{equation}
In an experiment, since generating many samples is impractical, one studies 
a single large sample with a particular realization of the quenched disorder
 of size $l$. For any $\xi$ the
 value of $X$ measured
in the sample
is a sampling from a probability distribution with relative variance 
$R_X(\xi,l)$. Thus $R_X(\xi,l)$ controls the deviation of $X$ from the many
 samples average.  If the system is governed by a random fixed point, 
 as the correlation length is increased, $R_X$ increases as 
$\sim \left(\frac{l}{\xi}\right)^{-d}$. 
$X$ behaves like the average of $\left(\frac{l}{\xi}\right)^{d}$ independent
measurements on regions
of size ${\xi}^{d}$. The variance of these measurements does not decrease as
 $\xi$ increases; it is constant.
On the other hand if the disorder is irrelevant
and the system is governed by a pure fixed point with $\alpha<0$,
 as the correlation length is increased, $R_X$ increases more mildly as 
$\sim \left(\frac{l}{\xi}\right)^{-d} \xi^{\frac{\alpha}{\nu}}$. In this case
 too, $X$ behaves like the average of $\left(\frac{l}{\xi}\right)^{d}$ 
measurements on regions
of size ${\xi}^{d}$. However, as $\xi$ increases, the variance of these 
measurements decreases as $\sim \xi^{\frac{\alpha}{\nu}}$. 


We have verified that for a disordered system governed by a random fixed point
$(\delta T_c(l))^2$ does not scale as $\sim l^{-d}$,
but rather that $(\delta T_c(l))^2 \sim l^{-\frac{2}{\nu}}$.
This is an important result, similar to the situation in the purely geometric
percolation problem\cite{Stauffer 1992}. Recently P\'{a}zm\'{a}ndi, Scalettar
 and Zim\'{a}nyi\cite{Pazmandi 1997}
claimed that the bound $\nu \geq 2/d$, which was supposed to hold
for disordered systems\cite{Chayes 1986}, is not valid. As they show,
if in systems violating this bound one would have 
$(\delta T_c(l))^2 \sim l^{-d}$ [our equation (\ref{eq: delta Tc})], then 
simulations at $T_c^\infty$ would not
be able to capture the true critical exponents\cite{Pazmandi 1997}.
In fact in \onlinecite{Pazmandi 1997} (\ref{eq: delta Tc}) is termed ``the
most likely scenario'' and the conclusion drawn is that
 ``self averaging breaks down''. 
However, studies of percolation\cite{Stauffer 1992}, our results, and those of 
AH\cite{Aharony 1996} imply the contrary.
 $(\delta T_c(l))^2 \sim l^{-\frac{2}{\nu}}$ [our equation
(\ref{eq: delta Tc AH})], and therefore simulations
at $T_c^\infty$ are able to capture the true critical exponents even if
$\nu < 2/d$. This also becomes evident by examining  the finite size scaling
 theory of \onlinecite{WD Self-Av} for $[X_i(T_c^\infty)]$, assuming that 
 (\ref{eq: delta Tc}) holds versus the consequences of (\ref{eq: delta Tc AH}).

We've shown that fluctuations in $X_i$ at $T_c^\infty$ are predominantly due
 to fluctuations in $\delta T_i= T_c^\infty-T_c(i,l)$, and that these fluctuations can be 
dramatically reduced by measuring $X_i$ at $T_c(i,l)$.
This suggests that using the histogram method to obtain $X_i(T_c(i,l))$ for
 each sample might be a better strategy for Monte Carlo studies than
the current strategy of studying $X_i(T_c^\infty)$.
It was also shown that to a good approximation, fluctuations of $X_i$
close to criticality can be accounted for by the finite size scaling form
(\ref{eq: disorder fss_noi}). We believe that a more extensive study of
the finite size scaling of sample to sample fluctuations is both feasible and
desired.


One of the surprising results of this work is the difference found between the
$p=0.6$ model with grand canonical disorder and the $c=0.6$ model
with canonical disorder. Our results indicate that for
$p=0.6$ and $c=0.6$ $V_{K_{\text{max}}}$ scales as $l^{-2y_t}$ and that 
$V_{K_{\text{max}}}(p=0.6)/ V_{K_{\text{max}}}(c=0.6) \approx 3.26$.
This is apparently the reason why for these two types of disorder
$R_X$ tends as $l \rightarrow\infty$ to different constants. On the other hand we did not find any
difference in the scaling exponents of the two types of disorder. 




\acknowledgments
We thank A. Aharony, K. Binder, A.B. Harris, W. Janke, M. Picco and D. Stauffer
  for useful discussions and suggestions,
H.O. Heuer for correspondence and D. Lellouch for advice on statistics.  
This research has been supported in part by the Germany-Israel Science
 Foundation (GIF) and in part by the Israel Ministry of Science.
Computations were performed in part on the Paragon at the HLRZ, Juelich, and in part on the SP2 at the Inter-University High Performance Computing Center,
 Tel Aviv.


\end{document}